\documentclass[12pt]{article}
\usepackage{a4p}
\usepackage{epsfig}
\usepackage{array}
\usepackage{cite}
\usepackage{pennames} 
\usepackage{rotating}
\usepackage{lscape}
\usepackage{hyperref}
\newcommand{\Opal}{\mbox{OPAL}}
\newcommand {\Z}  {\mathrm Z}

\newcommand {\epem} {\mathrm{e^+e^-}}
\newcommand {\x}    {{\mathrm X_0}}
\newcommand {\Ebeam} {E_{\mathrm{beam}}}
\newcommand {\Roff} {R_{\mathrm{off}}}
\newcommand {\sigmaa} {\sigma_{\mathrm{a}}}
\newcommand {\sigmamax} {\sigma_{\mathrm{max}}}
\newcommand {\deltaRres} {\delta R_{\mathrm{res}}}
\newcommand {\deltaZ} {\delta_{\mathrm{Z}}}
\newcommand {\RR} {R_{\mathrm{R}}}
\newcommand {\RL} {R_{\mathrm{L}}}
\newcommand {\phiR} {\phi_{\mathrm{R}}}
\newcommand {\phiL} {\phi_{\mathrm{L}}}
\newcommand {\ER} {E_{\mathrm{R}}}
\newcommand {\EL} {E_{\mathrm{L}}}
\newcommand {\Rin} {R_{\mathrm{in}}}
\newcommand {\Rout} {R_{\mathrm{out}}}
\newcommand {\cin} {c_{\mathrm{in}}}
\newcommand {\cout} {c_{\mathrm{out}}}
\newcommand {\tin} {t_{\mathrm{in}}}
\newcommand {\tout} {t_{\mathrm{out}}}
\newcommand {\Rmin} {R_{\mathrm{min}}}
\newcommand {\Rmax} {R_{\mathrm{max}}}
\newcommand {\Ndata} {N_{\mathrm{data}}}
\newcommand {\NMC} {N_{\mathrm{MC}}}
\newcommand {\NMCnorun} {N^0_{\mathrm{MC}}}
\newcommand {\deltaRextra} {\delta R_{\mathrm{extra}}}
\newcommand {\LEPone} {LEP\,1}
\newcommand {\LEPtwo} {LEP\,2}
\newcommand{\phz} {\phantom{0}}
\def\Was{W\c as}
\begin{document}
%
%
\begin{titlepage}
\begin{center}
\Large EUROPEAN ORGANISATION FOR NUCLEAR RESEARCH
\normalsize
\vspace{1cm}
\end{center}

\begin{flushright}
    \large
    CERN-PH-EP/2005-014 \\
    21 February 2005    \\
    Revised 28 June 2005
\end{flushright}
%
%
\begin{center}
    \huge\bf\boldmath
Measurement of the running of the QED coupling in small-angle Bhabha scattering
at LEP
\end{center}
\vspace{1cm}
\bigskip
%
%
\begin{center}
\LARGE
\Opal\ Collaboration \\
\normalsize
\vspace{1cm}
\bigskip
%
%
\begin{abstract}
\noindent
Using the OPAL detector at LEP, the running of the effective QED
coupling $\alpha(t)$ is measured for space-like momentum transfer 
from the angular distribution of small-angle Bhabha scattering. 
In an almost ideal QED framework, with very
favourable experimental conditions, we obtain:
\begin{displaymath}
\Delta\alpha(-6.07 \,\mathrm{GeV^2}) - \Delta\alpha(-1.81 \,\mathrm{GeV^2}) = 
(440 \pm 58 \pm 43 \pm 30) \times 10^{-5} \,,
\end{displaymath}
where the first error is statistical, the second is the experimental systematic
and the third is the theoretical uncertainty.
This agrees with current evaluations of $\alpha(t)$.
The null hypothesis that $\alpha$ remains
constant within the above interval of $-t$ is excluded with a
significance above $5\,\sigma$. 
Similarly, our results are inconsistent at
the level of $3\,\sigma$ with the hypothesis that only leptonic loops
contribute to the running.
This is currently the most significant direct measurement where the
running $\alpha(t)$ is probed differentially within the measured $t$ range.
\end{abstract}


\end{center}

%
\vspace{0.2cm}
\begin{center}
{\large Submitted to European Physical Journal C}
\end{center}

%
\end{titlepage}
\begin{center}{\Large        The OPAL Collaboration
}\end{center}\bigskip
\begin{center}{
G.\thinspace Abbiendi$^{  2}$,
C.\thinspace Ainsley$^{  5}$,
P.F.\thinspace {\AA}kesson$^{  3,  y}$,
G.\thinspace Alexander$^{ 22}$,
G.\thinspace Anagnostou$^{  1}$,
K.J.\thinspace Anderson$^{  9}$,
S.\thinspace Asai$^{ 23}$,
D.\thinspace Axen$^{ 27}$,
I.\thinspace Bailey$^{ 26}$,
E.\thinspace Barberio$^{  8,   p}$
% ...
}\end{center}\bigskip
\bigskip
$^{  1}$School of Physics and Astronomy, University of Birmingham,
Birmingham B15 2TT, UK
\newline
$^{  2}$Dipartimento di Fisica dell' Universit\`a di Bologna and INFN,
I-40126 Bologna, Italy
\newline
$^{  3}$Physikalisches Institut, Universit\"at Bonn,
D-53115 Bonn, Germany
\newline
% ...
$^{  5}$Cavendish Laboratory, Cambridge CB3 0HE, UK
\newline
% ...
$^{  8}$CERN, European Organisation for Nuclear Research,
CH-1211 Geneva 23, Switzerland
\newline
$^{  9}$Enrico Fermi Institute and Department of Physics,
University of Chicago, Chicago IL 60637, USA
\newline
% ...
$^{ 22}$Department of Physics and Astronomy, Tel Aviv University,
Tel Aviv 69978, Israel
\newline
$^{ 23}$International Centre for Elementary Particle Physics and
Department of Physics, University of Tokyo, Tokyo 113-0033, and
Kobe University, Kobe 657-8501, Japan
\newline
% ...
$^{ 26}$University of Victoria, Department of Physics, P O Box 3055,
Victoria BC V8W 3P6, Canada
\newline
$^{ 27}$University of British Columbia, Department of Physics,
Vancouver BC V6T 1Z1, Canada
\newline
% ...
$^{  p}$ now at The University of Melbourne, Victoria, Australia
\newline
% ...
$^{  y}$ now at CERN
\newline
% ...
\bigskip
\section{Introduction}
The electromagnetic coupling constant is a basic parameter
of the Standard Model, known
with a relative precision of $4 \times 10^{-9}$ \cite{codata98}
at zero momentum transfer.
In QED the effective coupling changes, or {\em runs},
with the scale of momentum transfer due to vacuum polarization. 
This is due to virtual lepton and quark loop corrections 
to the photon propagator. 
This effect can also be understood as an increasing penetration of the
polarized cloud of virtual particles which screen the bare electric
charge of a particle as it is probed at smaller and smaller distance.
The effective QED coupling is generally expressed as:
\begin{equation}
\alpha(q^2) = \frac{\alpha_0}{1-\Delta\alpha(q^2)} 
\end{equation}
where $\alpha_0 = \alpha(q^2=0) \simeq 1/137.036$ 
is the fine structure constant and $q^2$ is the momentum transfer
squared of the exchanged photon.

Whereas the leptonic contributions to $\Delta\alpha$ 
are calculable to very high accuracy in quantum electrodynamics, 
the hadronic ones are more problematic as they involve quark masses and
hadronic physics at low momentum scales.
The hadronic contribution can be evaluated through 
a dispersion integral over the measured cross section of
$\epem \rightarrow \mathrm{hadrons}$, supplemented with perturbative QCD 
at energies above the resonance region \cite{ej95,bp2001}. 
The main difficulty of this approach comes from the integration 
of experimental data in the region of hadronic resonances, 
which gives the dominant uncertainty on
$\Delta\alpha$.
The effective QED coupling $\alpha(q^2)$ is an essential ingredient for many
precision physics predictions. 
It contributes one of the dominant uncertainties
in the electroweak fits constraining the Higgs mass \cite{lepew04}.
There are also many evaluations which
are more theory-driven, extending the application of perturbative QCD down to
about 2~GeV, for example \cite{detroconiz}.
An alternative approach uses the Adler function \cite{adler}
and perturbative QCD in the negative $q^2$ ({\em space-like}) region 
\cite{jeger}, where $\Delta\alpha$ is a smooth function.

Precise measurements of the muon anomalous magnetic moment \cite{bailey,mug-2}
provide a powerful test of vacuum polarization corrections 
to the photon propagator,  
due not only to electrons but also to other leptons and hadronic states.
Due to their very high precision, such measurements currently
constitute a stringent overall test of the Standard Model,
in which many classes of radiative corrections are probed,
involving electromagnetic, strong and weak interactions. 
They also have sensitivity to potential physics beyond the Standard Model.
Vacuum polarization corrections affect the muon anomaly
when integrated over all possible values for the $q^2$ of the photon,
both positive and negative.
In contrast, in this analysis, 
we present a {\it direct} measurement of the running of the electromagnetic
coupling in the space-like region,
where by direct we mean that the change in strength of the effective
coupling is probed differentially as a function of $q^2$ of the
photon propagator.

There have been only a few direct observations of the running 
of the QED coupling \cite{topaz,opalaem,venus,l3,l3new}. Most of these analyses
involve measurements of cross sections and their ratios and obtain values of 
$\alpha(q^2)$ which are found to deviate from $\alpha_0$ or from the assumed
value of the coupling at some initial scale.
Theoretical uncertainties on the predicted absolute cross sections as well as
experimental scale errors 
can influence such determinations or reduce their significance. 
The TOPAZ \cite{topaz} and the OPAL
\cite{opalaem} experiments probed the running in the {\em time-like} region
(positive $q^2$) from $\epem$ annihilations to leptonic final states.
Far enough from the $\Z$ resonance these processes
are dominated by single photon exchange, although they substantially
involve the full electroweak theory.
Large-angle Bhabha scattering has been studied
by the VENUS \cite{venus} and L3 \cite{l3,l3new} experiments
to measure the running in the space-like region.
In this case both $s$- and $t$-channel 
$\gamma$-exchange diagrams are important and the
effective QED coupling appears as a function of $s$ or $t$ respectively.
Moreover, interference contributions due to $\Z$-exchange are also sizeable.

In this paper we measure the running of $\alpha$ in the space-like region,
by studying the angular dependence of small-angle Bhabha scattering 
using data collected by the OPAL detector at LEP.
Small-angle Bhabha scattering appears to be an ideal process
for a direct measurement of the running of $\alpha$ 
since the momentum transfer squared $t$ is simply and unambiguously 
related to the polar scattering angle, and the angular distribution
is modified by the running coupling which appears as $\alpha^2(t)$.
There has been only one similar previous study by the L3 experiment \cite{l3}.  

We confine ourselves to the small angular region used for the luminosity
measurement, which approximately corresponds to $2 \leq -t \leq 6$~GeV$^2$ 
at centre-of-mass energies near the $\Z$ resonance peak (\LEPone).
At this $t$ scale the average $\Delta\alpha$ is about 2\%. 
The number of small-angle Bhabha events is used to determine the
integrated luminosity, so that we will not make an absolute measurement of
$\alpha(t)$; rather we will look only at the 
variation of $\Delta\alpha$ across the acceptance, which is expected to be about
$0.5$\%, leading to a variation of the differential cross section of about 1\%.
An interesting property of this low $|t|$ region is that, although the absolute
$\Delta\alpha$ value is dominated by the leptonic contributions, 
the leptonic and hadronic components contribute about equally to its variation
across the region accessible to our measurements.

A crucial element in this work has been
the very high precision in measuring the scattering angle provided by the 
OPAL Silicon-Tungsten (SiW) luminometer \cite{lumipap}.
Among the experimental advantages are also 
the high available statistics and the purity of the data sample.
Not less important is the cleanliness of this kind of measurement 
from a theoretical point of view, 
as has been pointed out recently in \cite{arbuzov}.
Small-angle Bhabha scattering is strongly dominated by
single-photon $t$-channel exchange, while 
$s$-channel photon exchange is practically negligible.
The cross section is currently exactly calculable up to the leading 
${\cal O}(\alpha^2)$ terms in the QED photonic corrections 
(herein indicated as ${\cal O}(\alpha^2 L^2)$, 
where $L = \ln(|t|/m_\mathrm{e}^2)-1$ is the large logarithm). 
Many existing calculations are described 
in \cite{yellowrep} and were also widely cross-checked,
mainly to reduce the theoretical error on the 
determination of the luminosity at \LEPone.
Higher order terms are partially accounted for through exponentiation. 
A calculation accurate to the subleading ${\cal O}(\alpha^2)$
terms \cite{nllbha} also exists.
Corrections for $\Z$ interference are very small and well known, so that
small-angle Bhabha scattering near the $\Z$ pole can be considered an essentially
pure QED process. 
A comparison of data with such precise calculations can determine the value of
the effective QED coupling in the most accurate way without relying
on the correctness of the SU(2)$\times$U(1) electroweak model.

The paper is organized as follows: we explain the analysis method in
Section~\ref{sec:method},
the detector and its Monte
Carlo simulation are briefly described in Section~\ref{sec:technic} and
the event selection in Section~\ref{sec:selection}. 
The procedure to correct the data distributions is explained in 
Section~\ref{sec:anchoring} and checked in Section ~\ref{sec:fit}.
The experimental systematic uncertainties are described in
detail in Section~\ref{sec:syst}. 
The theoretical uncertainties are discussed in
Section~\ref{sec:theory}. The results are finally given in 
Section~\ref{sec:results}, and a concluding summary 
in Section~\ref{sec:conclusions}.

\section{Analysis method}
\label{sec:method}
The Bhabha differential cross section can be written in the following form
for small scattering angle:
\begin{equation}
\frac{\mathrm{d}\sigma}{\mathrm{d}t} = 
\frac{\mathrm{d}\sigma^{(0)}}{\mathrm{d}t}
{\left( \frac{\alpha(t)}{\alpha_0} \right) }^2 (1+\epsilon)~(1+\delta_{\gamma})
+ \deltaZ
\label{eq:xsec}
\end{equation}
where
\begin{equation}
\frac{\mathrm{d}\sigma^{(0)}}{\mathrm{d}t} = \frac{4 \pi \alpha_0^2}{t^2}
\label{eq:tdep}
\end{equation}
is the Born term for $t$-channel single photon exchange, $\alpha_0$ 
is the fine structure constant and $\alpha(t)$ is the effective coupling at
the momentum transfer scale $t$. Here $ \epsilon $ represents the
radiative corrections to the Born cross section, while
$\delta_{\gamma}$ and $\deltaZ$ are respectively the contributions of
photon and $\Z$ $s$-channel exchange, both dominated by interference with 
$t$-channel photon exchange. The
contributions of $\delta_{\gamma}$ and $\deltaZ$ are much smaller than those of
$ \epsilon $ and the vacuum polarization. 
Therefore, with a precise knowledge of
the radiative corrections ($\epsilon$ term) one can determine the effective
coupling $\alpha(t)$ by measuring the differential cross section.
The form of Equation~(\ref{eq:xsec}) is an approximation 
since the $\delta_\gamma$ term is not strictly factorizable
with the effective coupling $\alpha^2(t)$.
In fact the $s$-channel amplitude couples as $\alpha(s)$, where $s$ is the
centre-of-mass energy squared.
The practical validity of Equation~(\ref{eq:xsec}) is a consequence of
the smallness of the $\delta_{\gamma}$ term, which could even be neglected.

The counting rate of Bhabha events in the SiW luminometers is used to determine
the integrated luminosity, so that we cannot make an absolute measurement of 
$\alpha(t)$ without an independent determination of the luminosity.
Instead, 
the structure of the cross section as written in Equation~(\ref{eq:xsec}) 
easily allows the variation of $\alpha(t)$ over the accessible $t$ range to be
determined, since the dominant contribution to the cross section contains the factor
$(\alpha(t) / \alpha_0)^2$.  
At leading order the variable $t$ is simply related to the scattering angle,
$\theta$:
\begin{equation}
t = - s \, \frac{1- \cos\theta}{2} \approx - \frac{s \,\theta^2}{4} \,.
\label{eq:t}
\end{equation}
Photon radiation (in particular initial-state radiation) smears this
correspondence. 
The event selection that we use, described in Section~\ref{sec:selection}, 
has been carefully chosen to
reduce the impact of radiative events. In particular the energy cuts and the
acollinearity cut are very effective. As a result the event sample is strongly
dominated by two-cluster configurations, with almost full energy back-to-back
scattered $\mathrm e^+$ and $\mathrm e^-$. 
For such a selection Equation~(\ref{eq:t}) represents 
a good approximation.
The scattering angle is measured from the radial position $R$ of
the scattered $\mathrm e^+$ and $\mathrm e^-$ at reference planes 
located within the SiW luminometers, 
at a distance $z$ from the interaction point:
\begin{equation}
\tan \theta = R / z \,.
\label{eq:theta}
\end{equation}

We use the BHLUMI \cite{bhlumi} Monte Carlo generator 
for all calculations of small-angle Bhabha scattering. 
It is a multiphoton exponentiated generator accurate up to the 
leading logarithmic ${\cal O}(\alpha^2 L^2)$ terms. Higher order
photonic contributions are partially included by virtue of the exponentiation.
The generated events always contain the scattered
electron and positron plus an arbitrary number of (non-collinear) photons.
Small contributions from $s$-channel photon exchange and $\Z$ interference are also
included. Corrections due to vacuum polarization are implemented with a few
choices for the parameterization of the hadronic term \cite{ej95,bp95}.
We used the option to generate weighted events, such that we could access all
the available intermediate weights which contribute to the final complete 
cross section event by event. In particular we could also modify the
parameterization of the vacuum polarization or set $\alpha(t) \equiv \alpha_0$
to assume a fixed coupling $\alpha_0$.

We compare the radial distribution of the data (and hence the $t$-spectrum)
with the predictions of the BHLUMI Monte Carlo
to determine the running of $\alpha$ within the accepted region.
If the Monte Carlo is modified by setting the coupling to the constant value 
$\alpha(t) \equiv \alpha_0$, the ratio $f$
of the number of data to Monte Carlo events in a given radial bin is:
\begin{equation}
f(t) = \frac{N_{\mathrm{data}}(t)}{N_{\mathrm{MC}}^0(t)} \propto 
{ \left( \frac{1} {1-\Delta\alpha (t)} \right) }^2  \, .
\end{equation}
The dominant dependence of $\Delta\alpha (t)$ expected from theory is
logarithmic. 
We therefore fitted the ratio $f(t)$ as:
\begin{equation}
f(t) = a + b \, \ln \left( \frac{t}{t_0} \right)
\label{eq:Rfit}
\end{equation}
where $a$ and $b$ are the fit parameters and
$t_0 = -3.3$~GeV$^2$ is chosen to be close to
the mean value of $t$ in the data sample.
The value of $t$ in each bin is calculated according to Equation~(\ref{eq:t}),
averaged over the bin, assuming the cross section dependence of
Equation~(\ref{eq:tdep}).
The parameter $a$ ($a \approx 1$) is not relevant 
since the Monte Carlo is normalized to the data.
The slope $b$ represents the full observable effect of the running of
$\alpha(t)$, both the leptonic and hadronic components. 
It is related to the variation of the coupling by:
\begin{equation}
\Delta\alpha(t_2) - \Delta\alpha(t_1) \simeq 
\frac{b}{2} \,\ln \left( \frac{t_2}{t_1} \right)
\label{eq:beff}
\end{equation}
where $t_1$ and $t_2$ correspond to the acceptance limits.
The errors associated with these approximations are negligible, and
discussed in Section~\ref{sec:biases}.
The fit error on $b$ includes a small contribution from the correlated error on
$a$.

With the acceptance cuts specified in Section~\ref{sec:selection} and the
average centre-of-mass energy $\sqrt{s} = 91.2208$~GeV, the reference
$t$ range is: $t_1 = -1.81$~GeV$^2$, $t_2 = -6.07$~GeV$^2$,
$\ln (t_2/t_1) = 1.21$.
The expected value of the effective slope in this $t$ range is $b = 0.00761$, 
determined by using the Burkhardt-Pietrzyk parameterization \cite{bp2001}
of the hadronic contributions. 

It is important to realize which systematic
effects could mimic the expected running or disturb the measurement.
The most potentially harmful effects are biases in the radial coordinate. 
Most simply one could think of dividing the detector acceptance into two 
and determining the slope using only two bins.
In such a model the running is equivalent to a bias in the
central division of $70\,\mu$m. 
Biases on the inner or outer radial cut
have a little less importance and could mimic the full running for
$90$ or $210\,\mu$m systematic offsets respectively.
Concerning radial metrology, a uniform bias of $0.5\,$mm on all radii 
would give the same observable slope as the expected running.
Knowledge of the beam parameters, particularly the transverse offset and
the beam divergence, is also quite important.

\section{Detector, data samples and Monte Carlo simulation}
\label{sec:technic}
The OPAL detector and trigger have been described in detail elsewhere
\cite{opaldet}.
In particular this analysis is based on 
the silicon-tungsten luminometer (SiW), which was used
to determine the luminosity from the counting rate of accepted Bhabha events
from 1993 until the end of LEP running.
The SiW was designed to improve the precision of the 
luminosity measurement to better than $1$ per mille. In fact it achieved a
fractional experimental systematic error of $3.4 \times 10^{-4}$.
The detector and the luminosity measurement are extensively described in
\cite{lumipap}. Here we only briefly review the detector aspects 
relevant for this analysis. 

\begin{figure}
\begin{center}
\epsfxsize=0.95\textwidth
\epsfbox{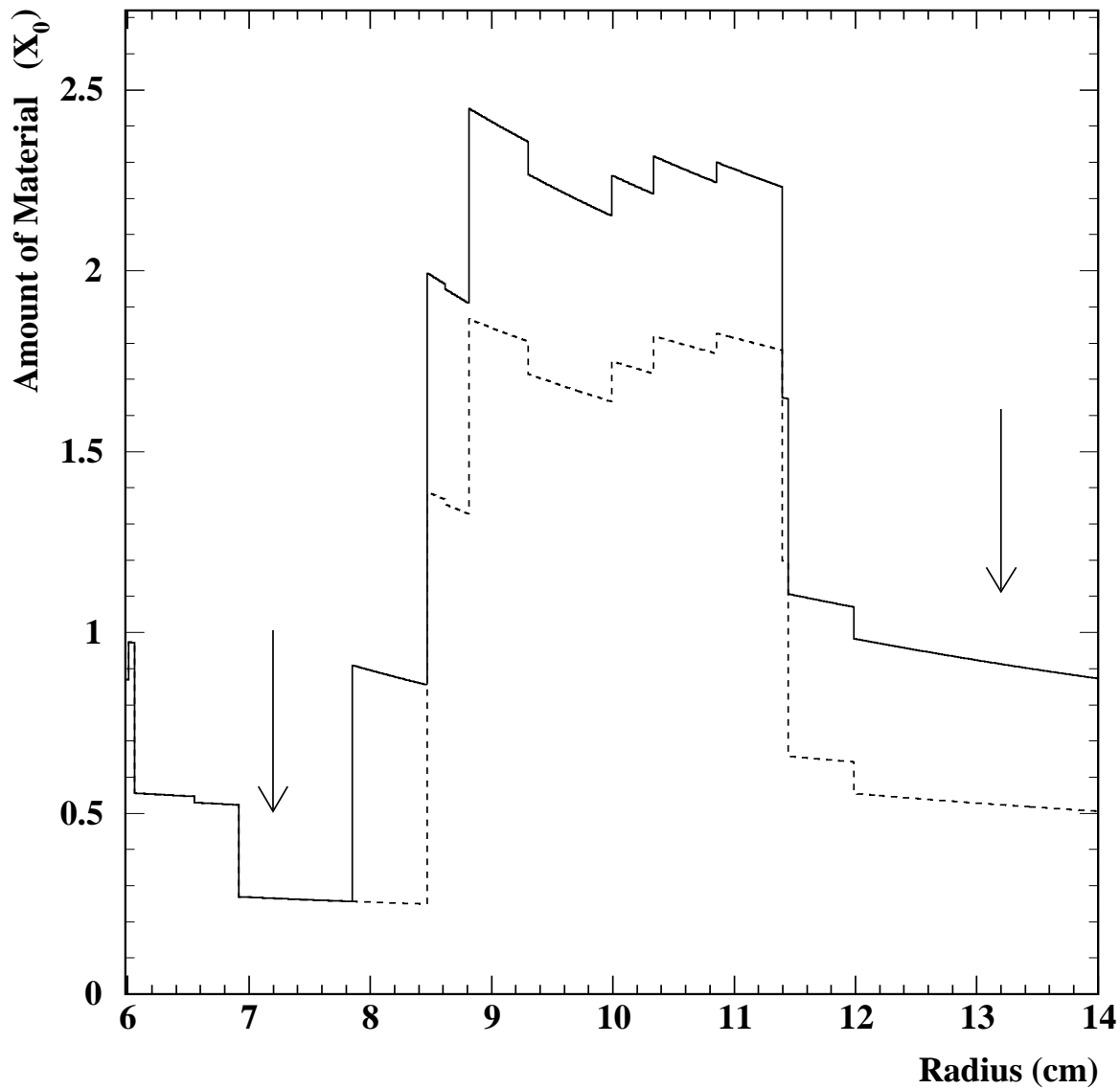}
\end{center}
\caption[Upstream material]{
The calculated thickness of material traversed by particles originating at
the interaction point as a function of calorimeter
radius for the 1993-94 detector configuration.
The solid curve corresponds to the Left, the dotted curve to the Right
side. The larger amount of material on the Left is
due to cables from the OPAL microvertex detector.
The arrows show the location of the 
acceptance definition cuts on shower radius.
}
\label{fig:dead-material-plot}
\end{figure}

The OPAL SiW luminometer consists of 2 identical cylindrical calorimeters,
encircling the beam pipe symmetrically at about $\pm 2.5$~m from the interaction
point. Each calorimeter is a stack of 19 silicon layers interleaved with 18
tungsten plates, with a sensitive depth of 14~cm, representing 22 radiation
lengths ($\x$). The first 14 tungsten plates are each $1\,\x$ thick, while the
last 4 are each $2\,\x$ thick. The sensitive area fully covers radii between 6.2
and 14.2~cm from the beam axis. 
Each detector layer is segmented with $R$-$\phi$ geometry in a 
$32 \times 32$ pad array. The pad size is 2.5~mm radially and 11.25 degrees in
azimuth. In total the whole luminometer has 38,912 readout channels
corresponding to the individual silicon pads. The calibration was studied with
electrical pulses generated both on the readout chips and on the front-end
boards, as well as with ionization signals generated in the silicon using test
beams and laboratory sources. 
Overall pad-to-pad gain variations were within 1\%.
Particles originating at the interaction point had to traverse the material
constituting the beam pipe and its support structures as well as cables from
inner detector components before reaching the face of the SiW calorimeters.
The distribution of this material upstream of the calorimeters is shown in
Fig.~\ref{fig:dead-material-plot}.
The material thickness was kept at a minimum especially in the crucial region
of the inner acceptance cut where it amounts to about $0.25\,\x$, 
while in the middle of the acceptance it increases to about $2\,\x$.
The model used for the calculation of the material thickness 
only approximates the actual situation, 
and dictates our strategy of determining the material 
effects from the data, using the figure as an approximate guide.
Controlling and correcting the possible biases in the reconstructed position
caused by this material was one of the most important aspects
of this analysis and is addressed in the following sections.

We use the data samples collected in 1993-95 at energies close to the $\Z$
resonance peak. 
In total they amount to 101~pb$^{-1}$ of OPAL data, 
corresponding to $12.0 \times 10^6$ accepted small-angle Bhabha events.
In 1993 and 1995, energy scans were performed and data were taken at three
energy points: close to the $\Z$ peak and approximately $1.8$~GeV above and
below it. In 1994 all data were taken near the $\Z$ peak. 
We divide the data into nine subsamples, depending on the year,
the centre-of-mass energy and the running conditions, such as
the beam parameters, in the same way as for the luminosity analysis
\cite{lumipap}. The same notation is also used for labelling the subsamples.
One subsample at peak energy (94-b) includes 40\% of the total number of 
events, while the remaining statistics are divided roughly equally
among the other, smaller, subsamples. 
When \LEPtwo\ data-taking started in 1996 the detector configuration changed,
with the installation of tungsten shields designed to protect the inner tracking
detectors from synchrotron radiation. These introduced about 50 radiation
lengths of material in front of the calorimeters between 26 and 33 mrad from the
beam axis, thus reducing the useful acceptance of the detector at the lower
polar angle limit. Moreover the new fiducial acceptance cut fell right in the
middle of the previous acceptance, where the preshowering material was maximum.
For these reasons we have limited this analysis to the \LEPone\ data samples.
 
The OPAL SiW detector simulation does not rely on a detailed physical simulation
of electromagnetic showers in the detector. Instead it is based on a
parameterization of the detector response obtained from the data \cite{lumipap}.
This approach
gives a much more reliable description of the tails of the detector response
functions, which are primarily due to extreme fluctuations in shower
development, than we could obtain using any existing program which attempts to
simulate the basic interactions of electrons and photons in matter. The measured
LEP beam size and divergence, as well as the measured offset and tilt of the
beam with respect to the calorimeters,
are also incorporated into this simulation. The Monte Carlo simulation 
is used to correct the acceptance for the effects of
the detector energy response, the coordinate resolution and LEP beam parameters.
Corresponding to each data subsample 
we generated an independent sample of BHLUMI events, 
using a slightly different set of parameters to match the experimental 
conditions in each case.
The statistics were always at
least 10 times those of the corresponding data set.

There are other effects which are not accounted for by the Monte
Carlo simulation, but rather studied directly in data. 
These include accidental background, detector metrology and, 
most importantly, biases in the reconstructed radial coordinate. 
The latter is crucial for
this analysis and will be discussed in Section~\ref{sec:anchoring}.

\section{Event selection}
\label{sec:selection}
The event selection criteria can be classified into {\em isolation} cuts, which
isolate a sample of pure Bhabha scattering events from the off-momentum
background, and acceptance defining, or {\em definition} cuts. The isolation
cuts are used to define a fiducial set of events which lie within the good
acceptance of both calorimeters and are essentially background free. The
definition cuts then select subsets of events from within the fiducial sample.

Showers generated by incident electrons and photons are recognized as clusters
in the calorimeters and their energies and coordinates determined. The fine
segmentation of the detectors allows incident particles with separations greater
than 1~cm to be individually reconstructed with good efficiency.

The coordinate system used throughout this paper is cylindrical, with the
$z$-axis pointing along the direction of the electron beam,
passing through the centres of the two calorimeter bores.
The origin of the azimuthal coordinate, $\phi$, is in the horizontal
plane, towards the inside of the LEP ring.
All radial coordinate measurements are projected onto reference planes 
at a distance of $\pm 246.0225$~cm from the nominal intersection point.
These reference planes correspond to the nominal position
of the silicon layers $7\,\x$ deep in the two calorimeters (hereafter we will
identify the relevant layers with expressions like
{\em layer}~7 or {\em layer}~$7\,\x$ without distinction).

The {\em isolation} cuts consist of the following requirements, imposed on
{($\RR$,$\phiR$)} and {($\RL$,$\phiL$)},
the radial and azimuthal coordinates of the highest 
energy cluster associated with the Bhabha event, 
in each of the Right and Left  
calorimeters, 
and on $\ER$ and $\EL$,
the total fiducial energy deposited by the Bhabha event
in each of the two calorimeters, explicitly including any detected energy
of radiated photons:
$$ 
\begin{array}{lrcl}
 
\bullet~ \mbox{Loose radial cut, Right (Left)} &
    6.7~{\rm cm} \  < \ \RR &<&13.7~{\rm cm} \\
&
    (6.7~{\rm cm}\  < \ \RL  &<&13.7~{\rm cm}) \\
~~~~~~~~~~~~~~~~~~~~~~~~~~~~~~~~~~~~~~~~~~~~~~~~~~~~~~
&  &&\\ 
\bullet~ \mbox{Acoplanarity cut} & 
    \left| \left|\phiR - \phiL \right| - \pi \right|&<&200~{\rm mrad} \\
&  &&\\ 
\bullet~ \mbox{Acollinearity cut} & 
    \left| \Delta R \right| = \left| \RR - \RL \right|&<&2.5~{\rm cm} \\
&  &&\\
\bullet~ \mbox{Minimum energy cut, Right (Left)} & 
    \ER &>&0.5\cdot \Ebeam \\
& 
   (\EL &>&0.5\cdot \Ebeam )\\
& &&  \\
\bullet~ \mbox{Average energy cut} & 
    \left( \ER + \EL \right)/{2} &>&0.75\cdot \Ebeam \\
\end{array}
 $$
Note that by defining the energy cuts relative to the beam energy,
$\Ebeam$, the selection efficiency is largely independent of $\sqrt{s}$.

The acollinearity cut (which corresponds to approximately
10.4~mrad) is introduced in order to ensure that the
acceptance for single radiative events is effectively determined
geometrically and not by the explicit energy cuts.

The isolation cuts accept events in which the radial coordinate, on both the
Right and the Left side, is more than two pad widths (0.5~cm) away from the edge
of the sensitive area of the detector.
The {\em definition} cuts, based solely on the reconstructed radial
positions $(\RR,\RL)$ of the two highest energy clusters, 
then require the radial position on one side or the other to be at least two
further pads towards the inside of the acceptance.
The Right and Left definition cuts are chosen so as to
correspond closely to radial pad boundaries in a given detector layer. 
When the chosen layer is the
reference layer at $7\,\x$, the definition cuts are:
$$
\begin{array}{lr}
\bullet~  \mathrm{Right~side} &
    7.2~{\rm cm}
    ~<~
    \RR
    ~<~
    13.2~{\rm cm} \\
~~~~~~~~~~~~~~~~~~~~~~~~~~~~~~&
~~~~~~~~~~~~~~~~~~~~~~~~~~~~~~~~~~~~~~~~~~~~~~~~~~~~~~~~~~~~~\\ 
\bullet~ \mathrm{Left~side} &
    7.2~{\rm cm}
    ~<~
    \RL
    ~<~
    13.2~{\rm cm} \\

\end{array} 
$$
Expressed in terms of polar angles, these cuts 
correspond to 29.257 and 53.602 mrad.

The measured radial distribution of Bhabha events 
will later be corrected with the procedure explained 
in Section~\ref{sec:anchoring}. 
This is based on one specific silicon layer, 
which can be varied with some freedom.
After a given layer is chosen, the
acceptance cuts and all the radial bin boundaries are defined according to
its pad boundaries, to match most closely with the correction procedure.
For example, when using the layer at $4\,\x$, the minimum and maximum radii
become $7.2584$~cm and $13.3071$~cm.

\begin{figure}
\begin{sideways}
\begin{minipage}[b]{\textheight}
\begin{center}
\begin{tabular}{cc}
\epsfxsize=0.46\textwidth\epsfbox[0 0 567 680]{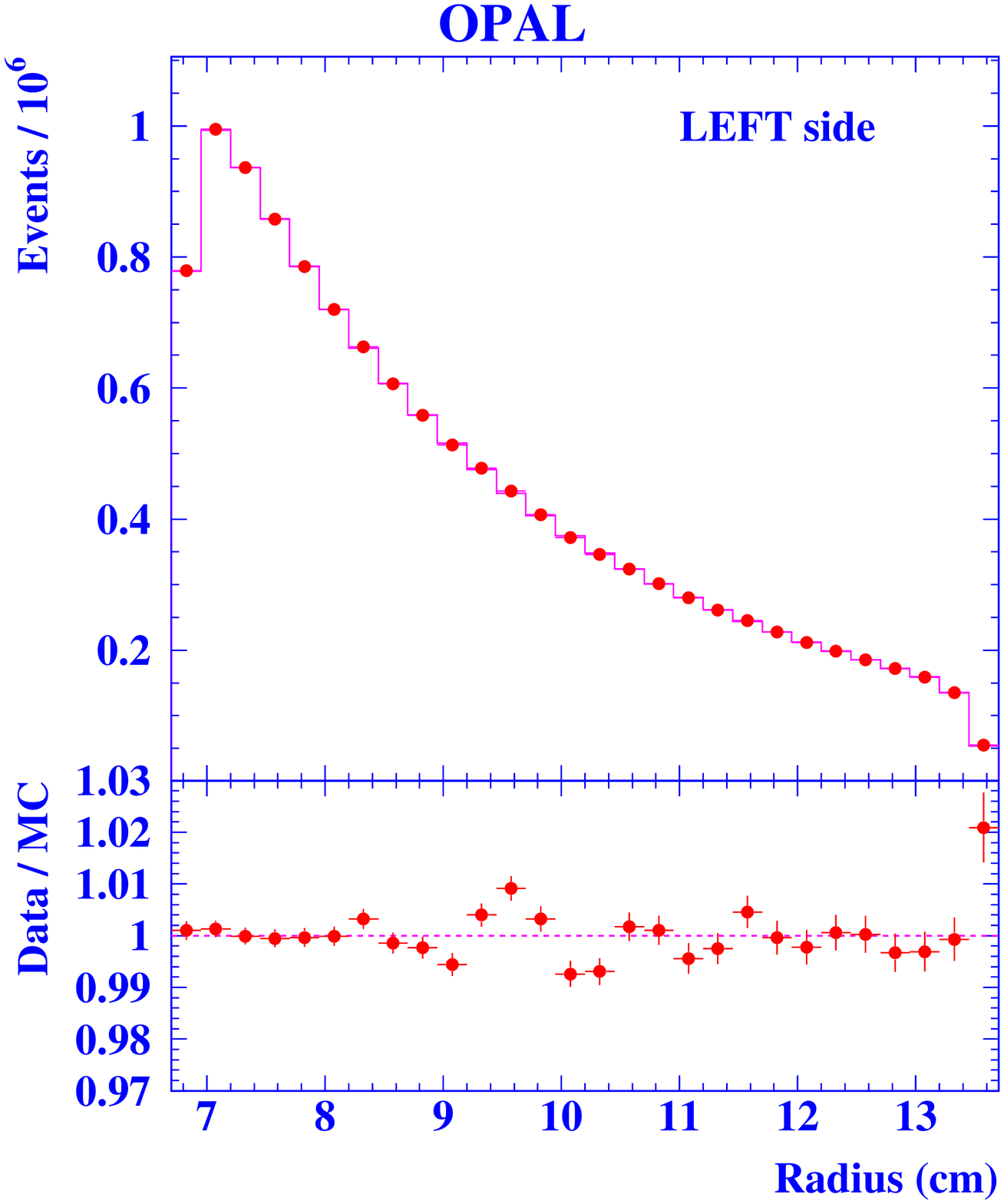}  &
\epsfxsize=0.46\textwidth\epsfbox[0 0 567 680]{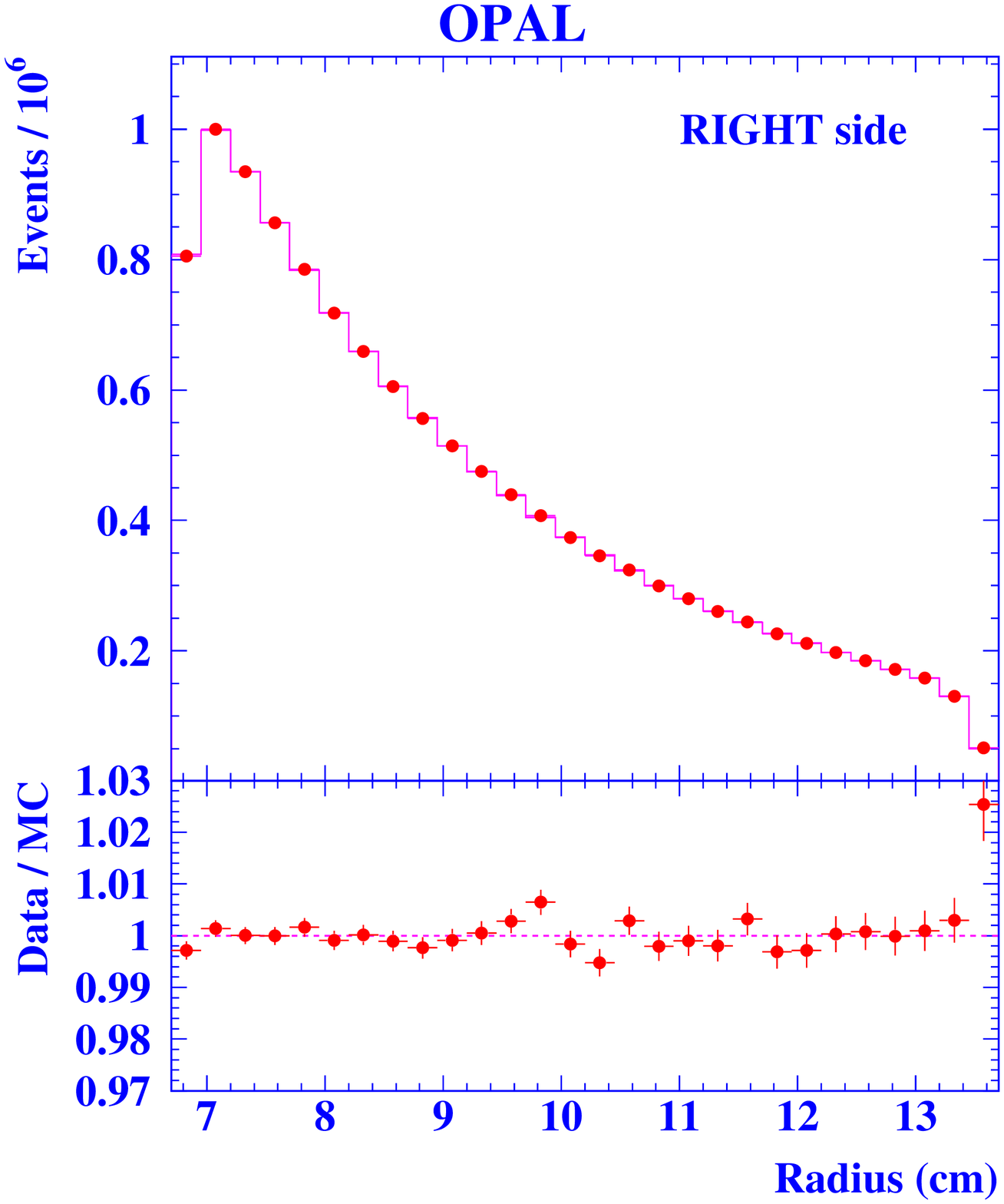}   \\
\end{tabular}
\end{center}
\caption
{Radial distributions after the isolation cuts for the complete data statistics
in the Left and Right calorimeters (mean $\sqrt{s}=91.2208$~GeV). 
The points show the data and the histogram
the Monte Carlo prediction, assuming the expected running of $\alpha$,
normalized to the same number of events.
The lower plots show the ratio between data and Monte Carlo.}
\label{fig:dradial}
\end{minipage}
\end{sideways}
\end{figure}

The radial distributions after the isolation cuts are shown in
Fig.~\ref{fig:dradial} for the complete data statistics and compared to Monte
Carlo distributions normalized to the same number of events. 
The agreement is good 
except in the central part, where effects from the preshowering material
are expected. Their correction is described in the following section.

The acceptance specified by the definition cuts is $0.5$~cm (corresponding to
two radial pad widths) wider than that used to define 
the OPAL luminosity \cite{lumipap}. 
In this way we extend the lever arm for observation of the running of $\alpha$. 
The compatibility of
the added data is quite good as can be seen from Fig.~\ref{fig:dradial}, where
each point corresponds to one pad width. The agreement has been quantified
by determining the $\chi^2$ increase obtained when the fit in the
former default acceptance is extended by 
2 or 3 pads at both the inner and outer radius. The $\Delta\chi^2$ is
consistent with purely statistical fluctuations corresponding 
to the added degrees of freedom. 

\section{Radial coordinate anchoring}
\label{sec:anchoring}
We exploit the fine radial and longitudinal granularity of the detector to
produce precise and continuous radial shower coordinates.
Limiting and quantifying the systematic error in these reconstructed
coordinates is crucial to the current measurement.
The key to ensuring that the reconstruction does not depart from the absolute
physical geometry of the Si pads, especially behind the appreciable preshowering
material which obscures the middle portion of the acceptance, is to study the
radial coordinate as a function of depth, using each of the many layers of the
detector independently. We refer to this procedure as {\em anchoring}.
Details of how the coordinates are formed from the recorded pad information are 
found in \cite{lumipap}, which should be consulted for a full understanding,
but the essentials are as follows. 
First the smoothed radial coordinate is constructed:
\begin {itemize}
\item
In each layer, the triplet of pads centred around 
the peak of the shower profile 
is used to form a continuous radial coordinate for the shower in 
that layer. 
The algorithm preserves two symmetry conditions, 
which hold to a good approximation. First, 
if the two largest pad signals in the triplet are equal, the coordinate falls
on the pad boundary. Secondly, if the signals on the two extreme pads of the 
triplet are equal, the coordinate falls at the centre of the central pad.
\item
Acceptable layer coordinates in layers 2 to 10 are projected onto the 
reference plane of the detector, layer 7, which lies near the average 
longitudinal shower maximum, and averaged.
\item
This average coordinate is then 
smoothed to compensate for non-uniformities in the resolution across the pad
structure of the detector.
The smoothing algorithm imposes the constraint that no 
coordinate is allowed to cross the radius of a pad boundary in the reference 
layer 7.
\end {itemize}
This smoothed radial coordinate will be referred to as the
{\em unanchored} coordinate.  This coordinate, derived from the pad signals
in many layers of the detector, is robust against fluctuations in any single
pad, and exhibits uniform, optimum, resolution.  In averaging many layers,
however, it does suffer from two inherent weaknesses: 
first, it does not preserve a
transparent relation to the absolute geometry of any particular pad.
Second, and even more important, it does not, in itself, allow the effect of the
preshowering material to be studied as a function of depth.

We therefore apply the {\em anchoring} procedure:
\begin {itemize}
\item
We estimate the residual bias, or {\em anchor}, of the smoothed radial
coordinate at each pad boundary in a given layer of the detector, as
explained in the following section. Here we rely on the simple symmetry
condition that two adjacent pads will each have a probability of 50\% for
recording the larger signal for an ensemble of events where
the shower axis lies on their common boundary. Having obtained the
apparent displacement of the pad boundaries through this condition, we
then obtain bin-by-bin acceptance corrections which will be applied to
the radial distribution. We then refer to the coordinate as being {\em
anchored}.
\end {itemize}

Note that in contrast to the unanchored coordinate, which relies on the
entire detector to provide the optimum position of an individual shower,
each anchor relies on the response of just two rings of radially adjacent pads, 
over the ensemble of showers which pass through them. 
Due to the cylindrical geometry of the detector, the radial coordinates of
the pad boundaries, projected to the reference layer~7, have a small
relative shift between adjacent layers of about $200~\mu$m at the inner
radius and about $350~\mu$m at the outer radius, and therefore provide a
dense net of benchmarks throughout the detector.

For the luminosity measurement, the anchoring procedure was essential in
establishing the absolute radius of the crucial inner acceptance boundary.
In the current analysis it becomes even more important, since it is used
to correct the acceptance for every bin of the radial distribution, as well
as to demonstrate that the effects of the preshowering material as a
function of depth are under control.
For this reason we discuss it here in some detail.

\subsection{Anchoring corrections}
\begin{figure}
\begin{center}
\epsfxsize=\textwidth
\epsfbox{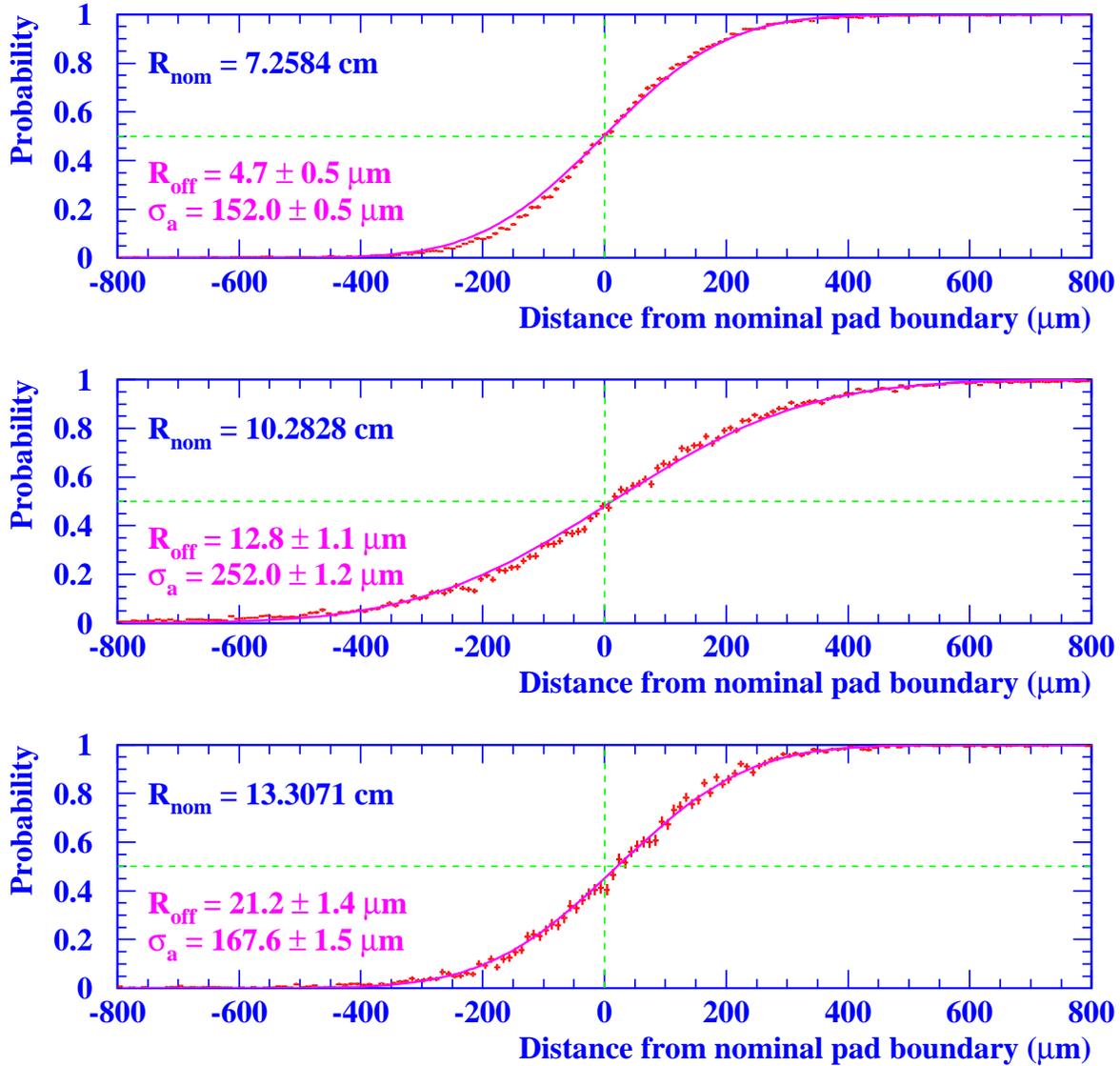}
\caption
{Pad boundary images obtained for the Si layer at $4\,\x$ at three different
radial positions corresponding to the inner edge, the middle portion 
and the outer edge of the acceptance. The solid curve shows the fitted
function, whose parameters $\Roff$ and $\sigmaa$ are given.}
\label{fig:pbimage}
\end{center}
\end{figure}

As the radial position of the incoming particles is
scanned across a radial pad boundary in a single layer, the probability for
observing the largest pad signal above or below this boundary shifts rapidly,
giving an image of the pad boundary as shown in Fig.~\ref{fig:pbimage}. 
These plots are obtained from OPAL data taken in 1993-94 and
refer to three radial pad boundaries in layer $4\,\x$ of the Right calorimeter.
Similar plots were also made for test beam data.
The pad boundary image is modelled with an error function: 
\begin{equation}
{\cal E}(R;\Roff,\sigmaa) = 
\int {\mathrm d}r \,{\cal G}(r;R,\sigmaa) \Theta(r - \Roff)
\end{equation}
where $R$ is the distance from the nominal pad boundary, 
${\cal G}(r;R,\sigmaa)$ is a Gaussian of width $\sigmaa$ and mean $R$, 
and $\Theta(r - \Roff)$ is a step function with offset $\Roff$ 
from the nominal pad boundary.  
The Gaussian width $\sigmaa$ 
measures the radial resolution at the boundary.
From Fig.~\ref{fig:pbimage} one can see that the width is similar at 
the inner and outer radius, while it is considerably greater at the central
radius.
The offset $\Roff$ is found to be quite small at the inner edge 
while it increases to
$\approx$\,10-20$\,\mu$m at the central and the outer radius. 

Despite our reliance on the 
basic symmetry condition of the reconstruction
described in Section~\ref{sec:anchoring}, 
in reality this symmetry is slightly violated due to the $R$-$\phi$ geometry 
of the pads.  
As a result, the mean position of an ensemble of showers which share energy
equally between two radially adjacent pads will actually lie at a smaller radius 
than the pad boundary between them.  The displacement between the point where
the energy, on average, is equally shared, and the actual pad boundary
is termed the {\em pad boundary bias},
$\delta R_{R\phi}$, and depends on the lateral extent of the shower.  
The pad boundary bias has been measured in a test beam and 
parameterized as a function of the apparent width of the shower, 
and varies from essentially zero to about $20\,\mu$m.

An additional, second order, effect also arises whenever a cut is imposed 
on a quantity with a steeply falling distribution, 
such as the radial Bhabha spectrum.
An acceptance change is introduced due to the fact that more events
actually on the uphill side of the cut will be measured to fall on the 
downhill side than vice versa.
This {\em resolution flow}, $\deltaRres$, is estimated to be
a small (positive) additional bias, typically below $1\,\mu$m and 
increasing to about $8\,\mu$m in the worst case.

The total net bias in the smoothed radial coordinate, $\delta R$ 
(also called anchor), is therefore given by:
\begin{equation}
\delta R = \Roff + \delta R_{R\phi} + \deltaRres
\label{eq:dr}
\end{equation}
where $\Roff$ is the observed offset of any particular pad boundary image, 
which may have positive or negative sign, while both $\delta R_{R\phi}$ and
$\deltaRres$ are always positive.

\begin{figure}
\begin{center}
\epsfxsize=\textwidth
\epsfbox{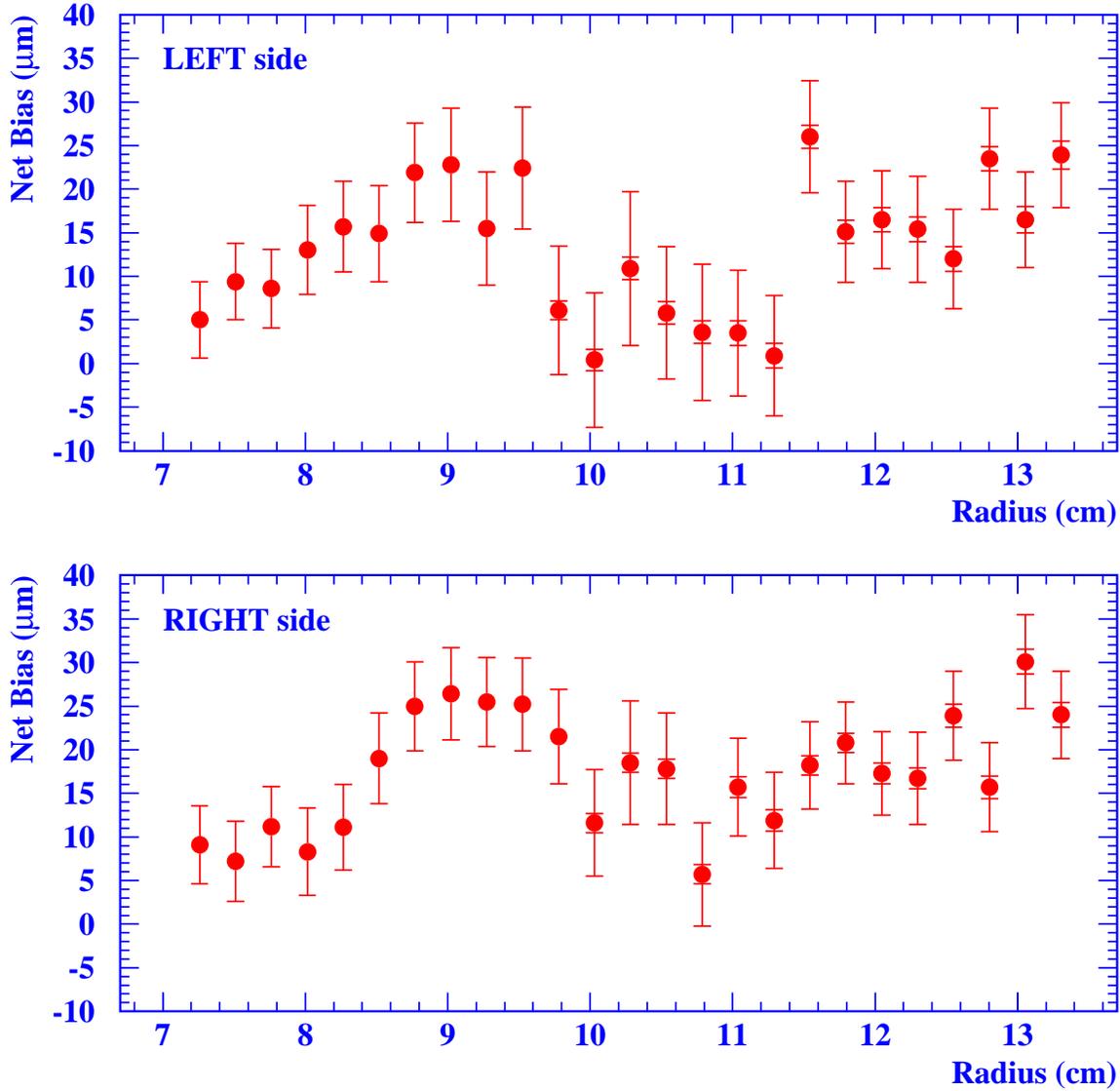}
\caption
{The total net bias $\delta R$ (anchor) as a function of radius for the Left
and the Right radial coordinate determined for the 
pad boundaries in the Si layer at $4\,\x$, for the combined 1993-94 data sample.
The full error bars show the 
total error, the inner bars the statistical component where this is larger
than the size of the points.}
\label{fig:anchors}
\end{center}
\end{figure}

The anchors determined from 1993-94 data for the layers at $4\,\x$ 
for all the pad boundaries used in the analysis are shown 
in Fig.~\ref{fig:anchors}. 
A similar trend is visible on the two sides, 
in particular the rise of the anchor from
$5$-$10 \,\mu$m at the inner edge to $20$-$25 \, \mu$m around $R=9$~cm.
The error bars include in quadrature the systematic errors 
from the fit method, pad gain variations, and the assumed $1/R$ scaling 
and shower width dependence of the pad boundary bias as discussed in
Section~\ref{sec:ancherr}.
The inner error bars show the statistical errors in the fit of the pad boundary
images. 
More details on errors assigned to the anchors can be found in \cite{lumipap}.
The anchors determined from 1995 data have similar features 
although with lower statistics.

\begin{figure}
\begin{center}
\epsfxsize=\textwidth
\epsfbox{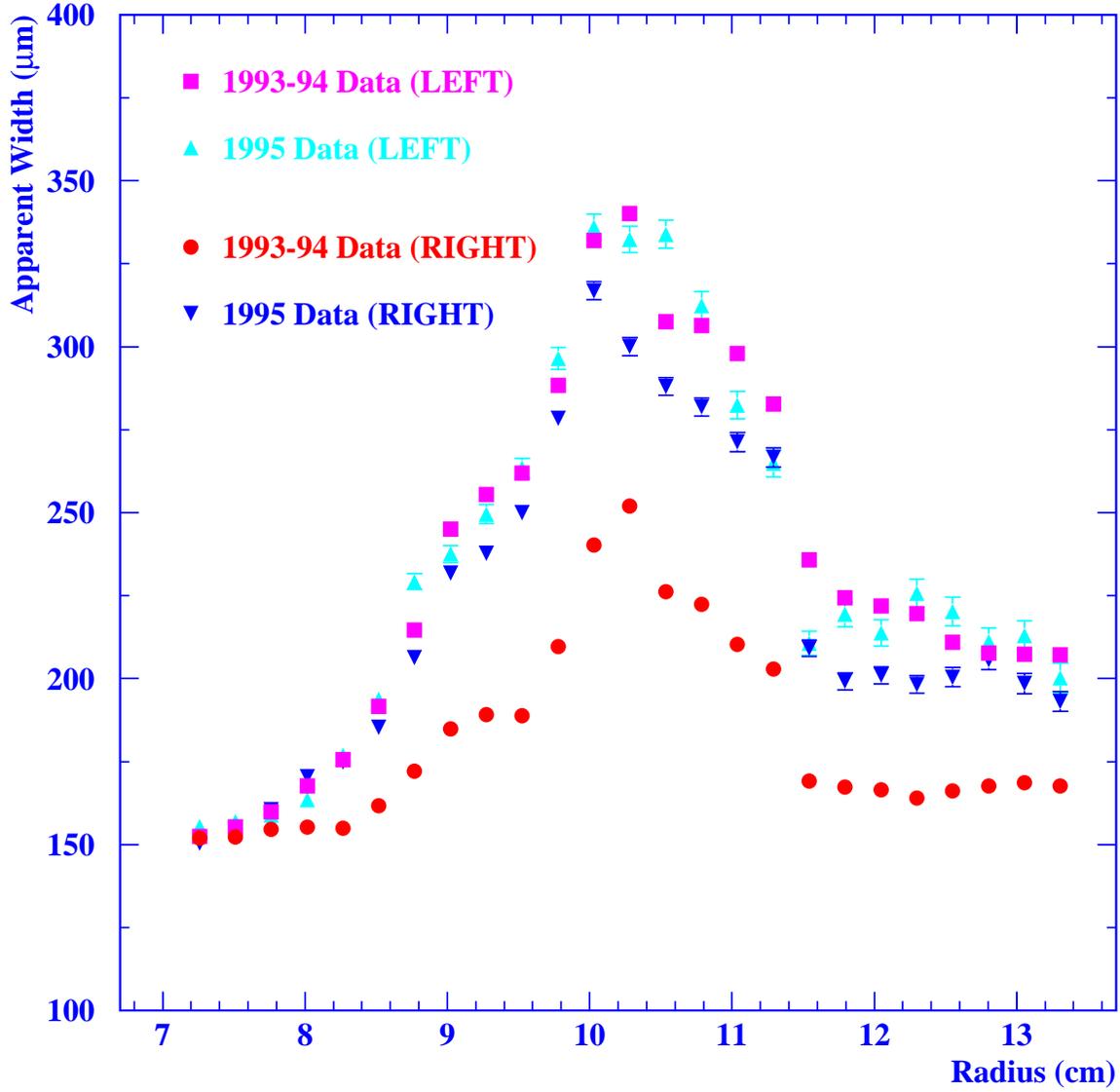}
\caption
{The apparent width $\sigmaa$ 
as a function of radius determined from the anchoring
procedure for the Si Layer at $4\,\x$ for
homogeneous data subsamples.
The errors are statistical only.}
\label{fig:widths}
\end{center}
\end{figure}

The anchors have been determined separately for the two homogeneous data
combinations, 1993-94 and 1995 data, 
because the amount of preshowering material was different 
in these two subsamples.
A clear relation with the amount and distribution of the material upstream of
the calorimeters is visible from the apparent width $\sigmaa$ as a function
of radius, as shown in Fig.~\ref{fig:widths}.
The noticeable difference between the Right and Left widths in
1993-94 data is due to the presence, on the Left,
of cables from the OPAL microvertex detector.
For 1995 data additional cables were installed on the Right side,
which resulted in an almost symmetrical situation.
The presence of a non-negligible amount of preshowering material in the middle
of the acceptance constitutes the most delicate 
experimental problem for this analysis, since the anchoring procedure
was developed and checked using the test beam only for the amount of
preshowering material ($< 1\,\x$) most relevant for the luminosity
measurement.
We have therefore made extensive checks as described 
in Section~\ref{sec:anch_lim} and~\ref{sec:fit} 
to identify a broad region within the full depth of the detector where
the anchoring procedure is valid, and derive our results using only 
the middle of this region.

To use the anchors, we do not actually correct the measured shower coordinates
themselves, but rather convert the anchors to appropriate acceptance corrections.
The acceptance of an individual radial bin with boundaries ($\Rin,~\Rout$)
is affected by the net biases of the edges $\delta \Rin,~\delta \Rout$ 
determined as in Equation~(\ref{eq:dr}) according to the following formula,
which gives the fractional acceptance variation:
\begin{equation}
\frac{\delta A}{A} = \cin \ \delta \Rin - \cout \ \delta \Rout
\label{eq:anc1}
\end{equation}
The coefficients $\cin$ and $\cout$ are 
derived by a simple analytical calculation assuming a
$1/\theta^3$ spectrum for the angular distribution and are given by:
\begin{equation}
c_{k} = 2\,\frac{\Rin^2 \Rout^2}{(\Rout^2 - \Rin^2)\,R_k^3} 
\hspace{1cm} k = \mathrm{in,out}
\label{eq:anc2}
\end{equation}
The radial distribution is binned according to the nominal pad boundaries of the
anchoring layer, projected onto the reference layer $7 \,\x$.
These anchoring corrections for a single 2.5~mm bin are
at most 0.5\% for the Right and 1.0\% for the Left side
in 1993-94 data and correspondingly 0.8\% and 0.7\% in 1995 data.

\subsection{Limitations in the anchoring}
\label{sec:anch_lim}
Our best security against the presence of excess systematic error in
the radial shower coordinate is to check for consistency between
the anchored and unanchored coordinates.
As already mentioned, the unanchored coordinates are robust, and derived from
the signals observed on a large number of pads throughout the detector, while
the anchored coordinates rest on observing equal signals, in the mean, 
inside and outside
an individual radial pad boundary in a particular detector layer.
These anchored coordinates provide reliable benchmarks throughout most of
the detector, but are expected to become fragile at both very shallow and
very deep layers of the calorimeters, particularly in the region of
the detector obscured by significant preshowering material.
Not only does the lateral shower profile broaden deep within the calorimeter, 
but beyond shower maximum the energy in the shower also becomes smaller.
Both these effects make determination of the pad boundary transition 
increasingly subject to disturbance.  
At shallow depths, particularly behind preshowering material
located considerably upstream of the detector, 
a shower can occasionally develop a long,
asymmetric tail which can lead to a significantly non-Gaussian error 
in the position determined in a single layer.

The pad boundary bias was determined in the test beam behind a maximum
of $0.84\,\x$ of preshowering material.  Under such conditions, and at
reasonable depths, we determined that the pad boundary bias could be
adequately parameterized as a function of the apparent shower width
alone.  Behind greater amounts of preshowering material, the validity of
this simplification may break down, and we can expect that an
additional, explicit, depth dependence may become necessary to
adequately describe the pad boundary bias at both very shallow and very
deep layers within the calorimeter.

We therefore compared the anchored and unanchored coordinates as a
function of the layer used for the anchoring to expose such effects.  As
expected, significant deviations are observed at both very shallow and
very deep layers, particularly on the Left side in the 1993-94 data,
where the preshowering material was greatest.

\begin{landscape}
\begin{figure}
\begin{center}
\epsfysize=0.9\textheight
\vspace{-1.5cm}
\epsfbox[0 0 850 567]{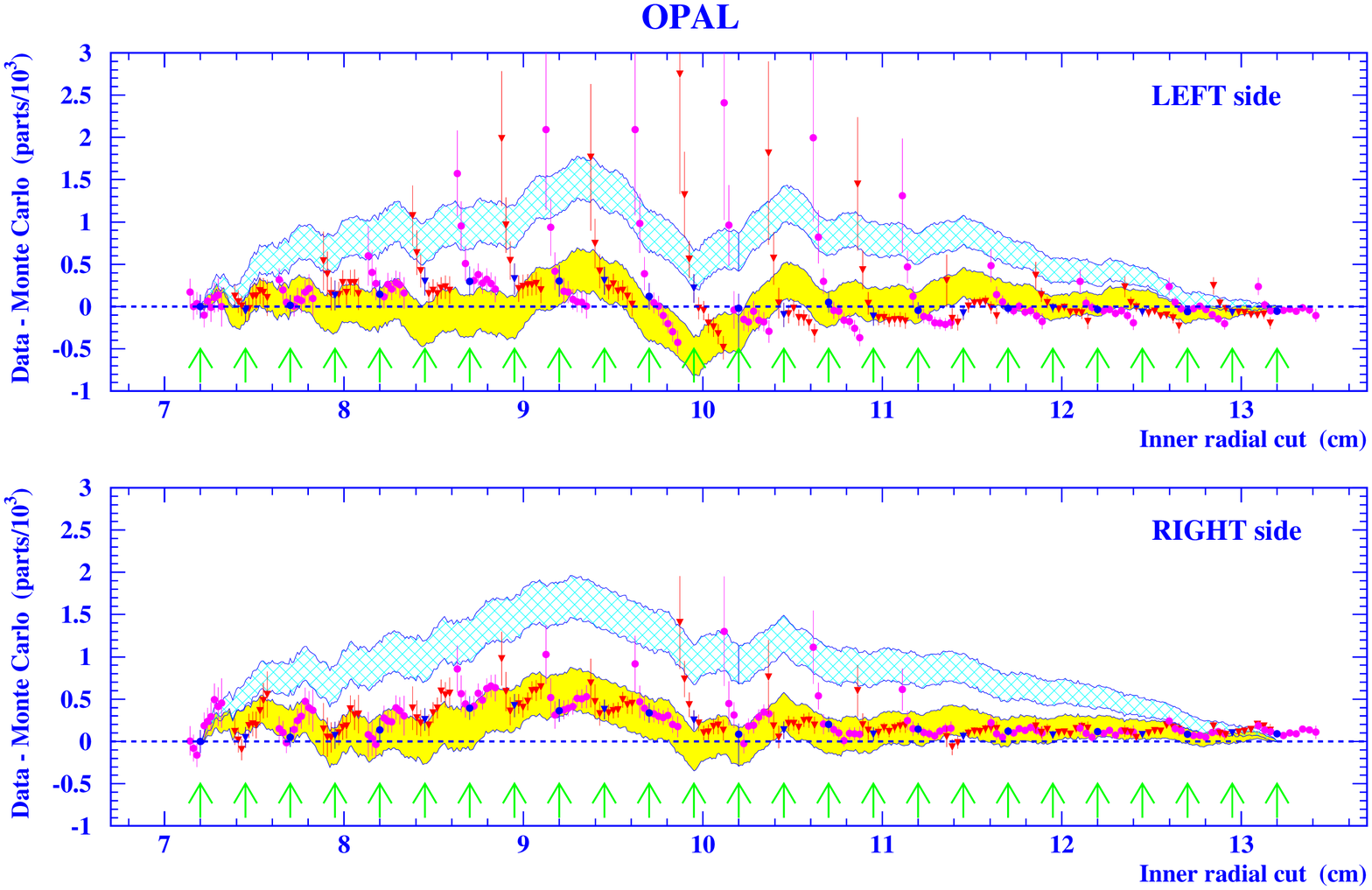}
\vspace{-0.8cm}
\caption
{Study of the reconstructed radial coordinate.
The shaded band shows the relative change in acceptance between data and Monte
Carlo as a function of the definition cut on the inner radius for the combined
1993-94 data sample. The width of the band shows the (highly correlated)
statistical errors.
The solid points show the acceptance variation determined
by the anchoring procedure in Si layers from $1\,\x$ to $10\,\x$, 
with the total errors shown.
Triangles and circles designate even and odd rings of radial pads respectively.
Within each group of points
the arrow shows the location of the radial pad boundary in layer $7\,\x$,
with deeper layers to the left and shallower layers to the right.
The anchor at $R=7.2$~cm is fixed to lie at zero.
The hatched band shows the prediction for zero running 
($\alpha(t) \equiv \alpha_0$).
}
\label{fig:dickplot}
\end{center}
\end{figure}
\end{landscape}

The reconstructed radial coordinate can be studied by 
comparing data with Monte Carlo as the radial acceptance cut is varied.
The Monte Carlo here assumes the expected running of $\alpha$ and that the
radial coordinate is reconstructed without bias.  
Thus differences in the acceptance of the data and Monte Carlo as
the radial cut is varied, beyond those expected from the finite statistics
and any departure from the expected running of $\alpha$,
can indicate biases in the radial coordinate.
We indicate with $N(R_1 ; R_2)$ the number of events falling 
between radii $R_1$ and $R_2$.
The acceptance change  
obtained by moving the inner radial cut to any given position $R$
is then proportional to $\Delta N = N(R;\Rmax) - N(\Rmin;\Rmax)$,
where $\Rmin=7.2$~cm and $\Rmax=13.2$~cm are the acceptance limits.
We normalize the Monte Carlo to the data, 
$\NMC(\Rmin;\Rmax) = \Ndata(\Rmin;\Rmax)$ and then form the quantity:
\begin{equation}
\left(\frac{\Delta A}{A}\right)_{\mathrm{data}} 
- \,\left(\frac{\Delta A}{A}\right)_{\mathrm{MC}} = 
\frac{\Ndata(R;\Rmax) - \NMC(R;\Rmax)}{\Ndata(\Rmin;\Rmax)} . 
\end{equation}
This relative acceptance, as a function of $R$, 
is shown as a shaded band (the lower one) 
for the Right and the Left side 
selection in Fig.~\ref{fig:dickplot} for 1993-94 data.
The width of each band represents
the binomial errors with respect to the reference selection
$7.2$~cm~$\leq R \leq 13.2$~cm. 
Note that, by construction, both ends of the relative acceptance band at
$R=7.2$~cm and $R=13.2$~cm lie at zero.

The solid points show the anchoring results for all the relevant pad boundaries
in layers between $1\,\x$ and $10\,\x$. 
The radial bias corresponding to each anchor is converted into an
acceptance variation using the formula:
\begin{equation}
\frac{\delta A}{A} = 2~\frac{\Rmin^2 \Rmax^2}{\Rmax^2 - \Rmin^2} ~
\frac{\delta R}{R^3}
\label{eq:dickformula}
\end{equation}
Since the normalization is the total acceptance, the low $R$ points
have a greater influence in the plot, due to the $1/R^3$ weighting. 
Therefore any visible structure tends to be flattened at increasing radius. 
Any of the anchors can be chosen to fix the absolute offset 
in the continuous radial coordinate.
Here we choose the anchor at $R=7.2$~cm in layer $7\,\x$ and this point
correspondingly lies at zero. 

Each group of nearby points, marked by either circles or triangles, 
refers to a given radial pad boundary, 
and the individual points in each group to different layers, 
at variable depth in the calorimeters.
Due to the origin of the projection at the beam interaction point, the radial
coordinates of these boundaries, projected to the reference layer~7, have a
small relative shift between adjacent layers 
of about $200~\mu$m at the inner radius and 
about $350~\mu$m at the outer radius.
The arrows mark the position of a
given pad boundary in layer $7\,\x$, deeper layers have a lower $R$ and
shallower layers a higher $R$.
Note that in contrast to the acceptance bands also shown in this figure, 
these anchoring points are independent of the Monte Carlo, 
and do not depend on the assumed running of $\alpha$.

While the broad features of the acceptance bands are sensitive to the
running assumed in the Monte Carlo, the small-scale, local structure in the
bands reveals even very small residual imperfections in the reconstructed
coordinate.
The essential point of Fig.~\ref{fig:dickplot}
is that where the anchor points derived from
individual pad boundaries follow these local variations in the relative
acceptance band the radial bias determined by the anchoring procedure
is seen to be consistent with this residual structure in the unanchored 
coordinate.
In contrast, wherever the anchor points exhibit a local pattern of divergence
from the acceptance band, especially as a function of depth, this indicates
an anomaly in these anchors.
Most anchors are found to be consistent with the relative acceptance band.
However, clear discrepancies are apparent 
for the deepest layers considered ($8$-$10 \,\x$), in particular for the Left side. 
This is most evident in the central region of acceptance, where
the amount of material between the detector and the
interaction point is large and the parameterization
of the expected bias derived from the test beam is evidently 
no longer applicable.
The behaviour of the anchors with depth indicates that the onset of these
problems is abrupt, and a large region of the detector remains well
understood for use in our analysis.
Notice that in the obscured region of the detector the run-away of the
anchors occurs almost exactly one layer earlier on the Left side of the
detector, at layer $8\,\x$, than it does on the Right, 
indicating that the $0.5\,\x$ of additional preshowering material 
has an effect larger than an equivalent amount
of compact absorbing material within the detector.

The plotted relative acceptance band depends on the running of
$\alpha$ that we want to measure, so its remarkably flat shape
means that the data agree with the input $\alpha(t)$ in the Monte
Carlo. To make this clear we have also plotted the prediction for zero running
as the hatched bands. 
The highly significant ``eyebrow'' shape of the zero-running
acceptance bands and their clear separation from the flat Standard Model bands
is a graphic representation of the sensitivity of our measurement.

The run-away anchors in the deepest layers considered are consistent 
with the eyebrows, however, and show
that the effects of the preshowering material for these deep layers, 
particularly on the Left side, 
would give an apparent shape consistent with zero running.
As mentioned previously this is equivalent to biases of about $70 \, \mu$m 
in the middle of the acceptance.

\section{Finding safe anchors for the measurement}
\label{sec:fit}
In order to see more directly the effect of anchoring imperfections
on the fitted slope $b$, 
we have made a series of test fits for each choice of anchoring layer.
The fits are simple $\chi^2$ fits of the ratio of data to Monte Carlo events
observed in each bin to the two coefficients of Equation~(\ref{eq:Rfit}).
The Monte Carlo in this case assumes a constant coupling 
$\alpha(t) \equiv \alpha_0$.
Since anchoring problems may manifest themselves either as increased
fluctuations from bin to bin, or as broader, more dangerous, systematic
effects highly correlated between nearby bins, we pursue two lines of
investigation.

\begin{figure}
\begin{center}
\epsfxsize=\textwidth
\epsfbox{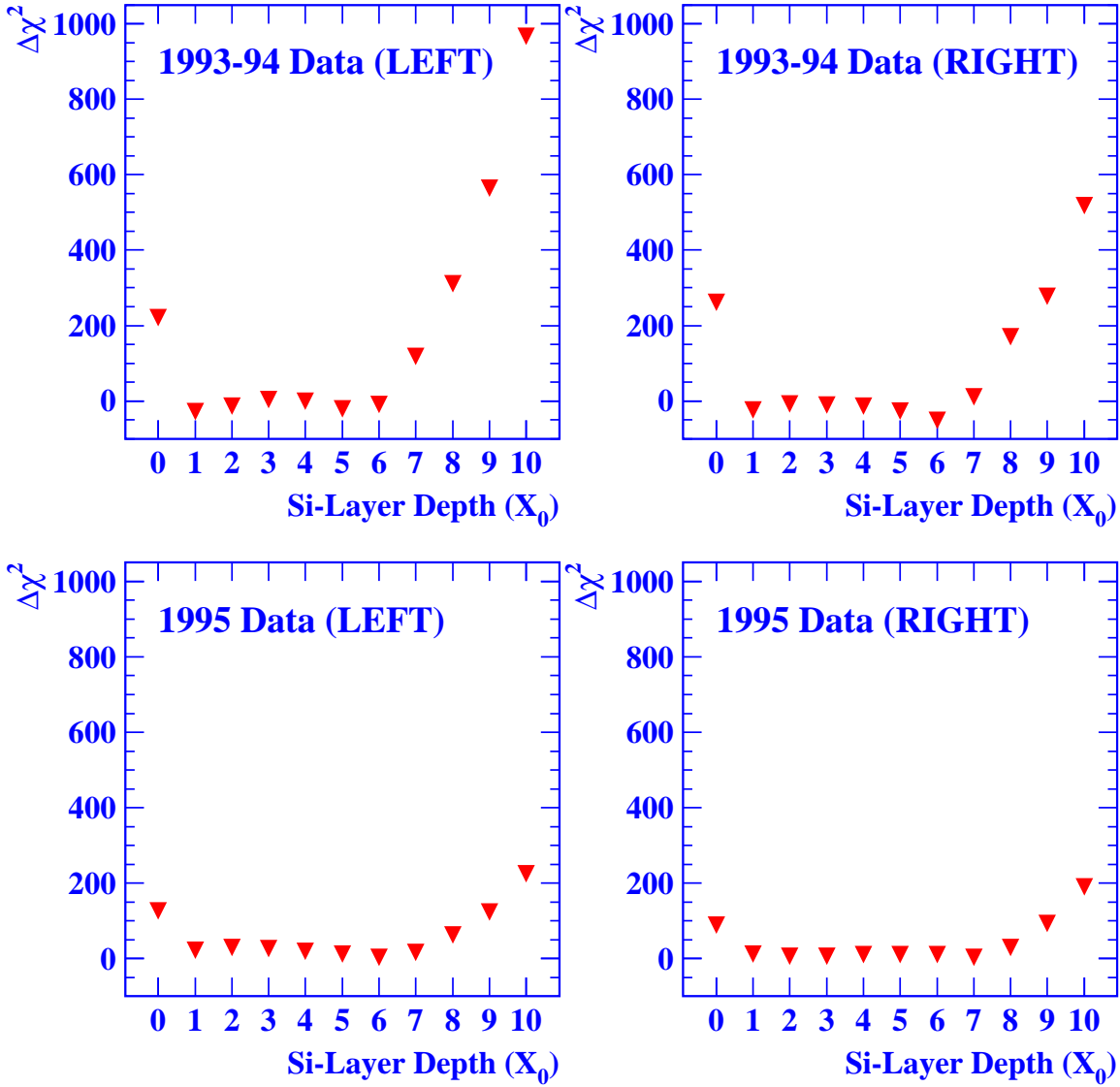}
\caption
{Difference in $\chi^2$ (for 22 degrees of freedom) between the fits of 
anchored and unanchored radial distributions 
as a function of the anchoring layer 
for homogeneous data subsamples.
Only statistical errors are considered.}
\label{fig:dchi2}
\end{center}
\end{figure}

In the first 
we divide the radial distribution into the maximum number
of bins: at each layer a bin corresponds to one of the 24 pads in the 
fiducial region.
Since the assessed systematic errors are somewhat correlated from bin to bin,
we consider only statistical errors, which are in any case dominant.
We focus on the difference in $\chi^2$
between the anchored and unanchored distributions, which is shown in
Fig.~\ref{fig:dchi2} as a function of the anchoring layer.
The results are shown separately for the two sides, 
and for the two homogeneous data combinations, 
1993-94 and 1995.
It is apparent that beyond layer $7\,\x$
the $\chi^2$ becomes progressively worse.
This agrees with the run-away of the anchors in deep layers
which was shown in Fig.~\ref{fig:dickplot}.
In contrast, the anchoring procedure appears stable in layers $1$-$6\,\x$.

\begin{figure}
\begin{center}
\epsfxsize=\textwidth
\epsfbox{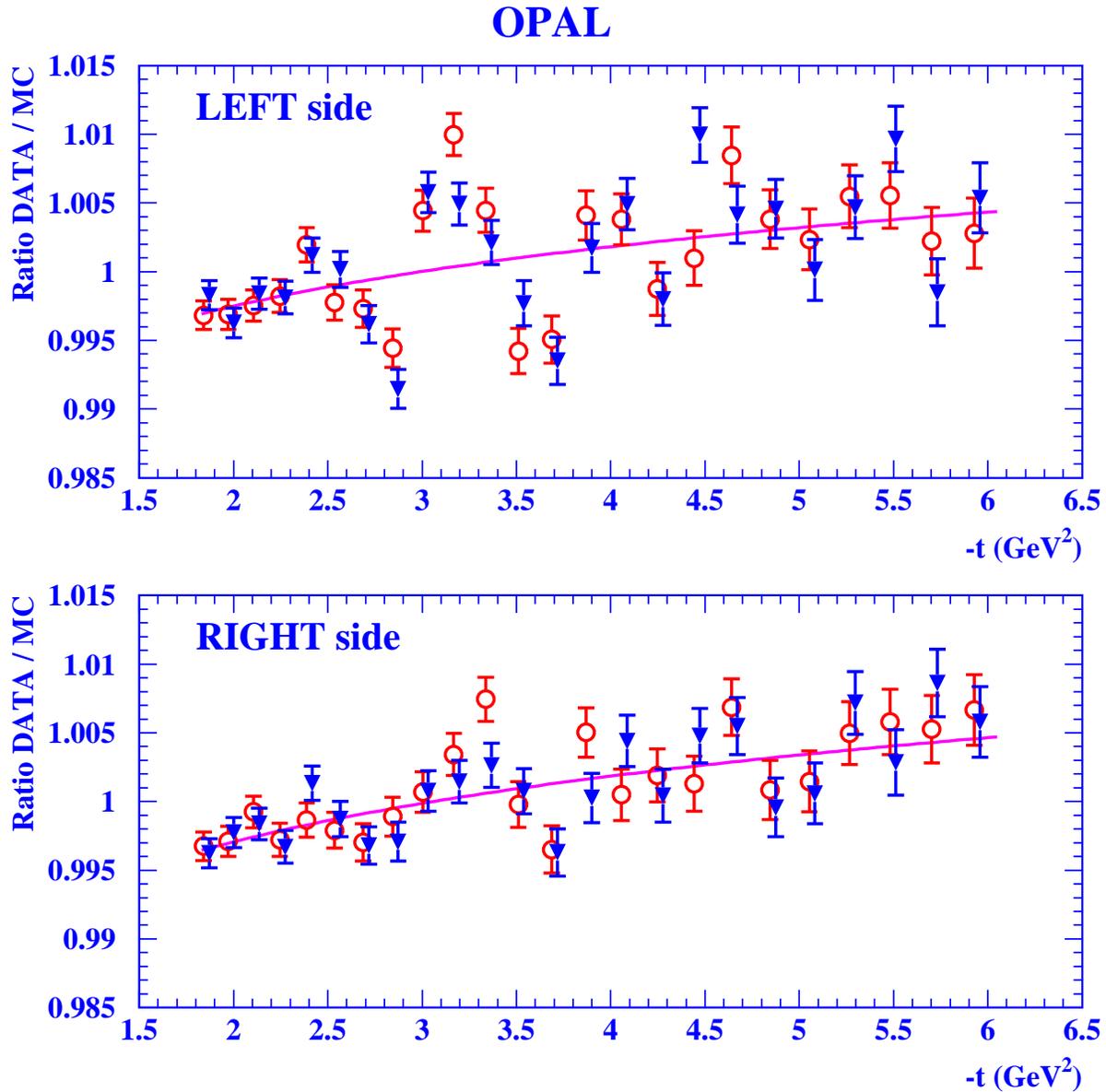}
\caption
{Ratios of numbers of data and Monte Carlo events 
(with $\Delta\alpha$ set to zero) for the
complete data statistics, for the Left and Right sides.
Each point corresponds to a bin of
one radial pad. The solid triangles show the data corrected with anchors in
layer $4\,\x$, the empty circles the unanchored data 
(slightly shifted horizontally for clarity), 
with statistical errors in both cases. The line shows the fit 
to the anchored data.}
\label{fig:fit.naive.comb}
\end{center}
\end{figure}

Fig.~\ref{fig:fit.naive.comb} gives an illustration for the combination 
of all data, anchored at layer $4\,\x$.
The fit $\chi^2$ is $41.3$/$22$ for the Right side and $133.2$/$22$ for the Left
side after the anchoring correction. 
The fit quality is much worse for the Left side
as is apparent from the greater dispersion of the points. 
The anchoring procedure improves the $\chi^2$ for the Right side 
but not for the Left: unanchored distributions have 
$\chi^2 = 47.9$/$22$ and $124.6$/$22$ respectively.
Allowing for the known anchoring systematic errors, as explained in 
Section~\ref{sec:ancherr}, a good $\chi^2$ is obtained for the 
Right side distribution, while the $\chi^2$ on the Left is still too large.

\begin{figure}
\begin{center}
\epsfxsize=\textwidth
\epsfbox{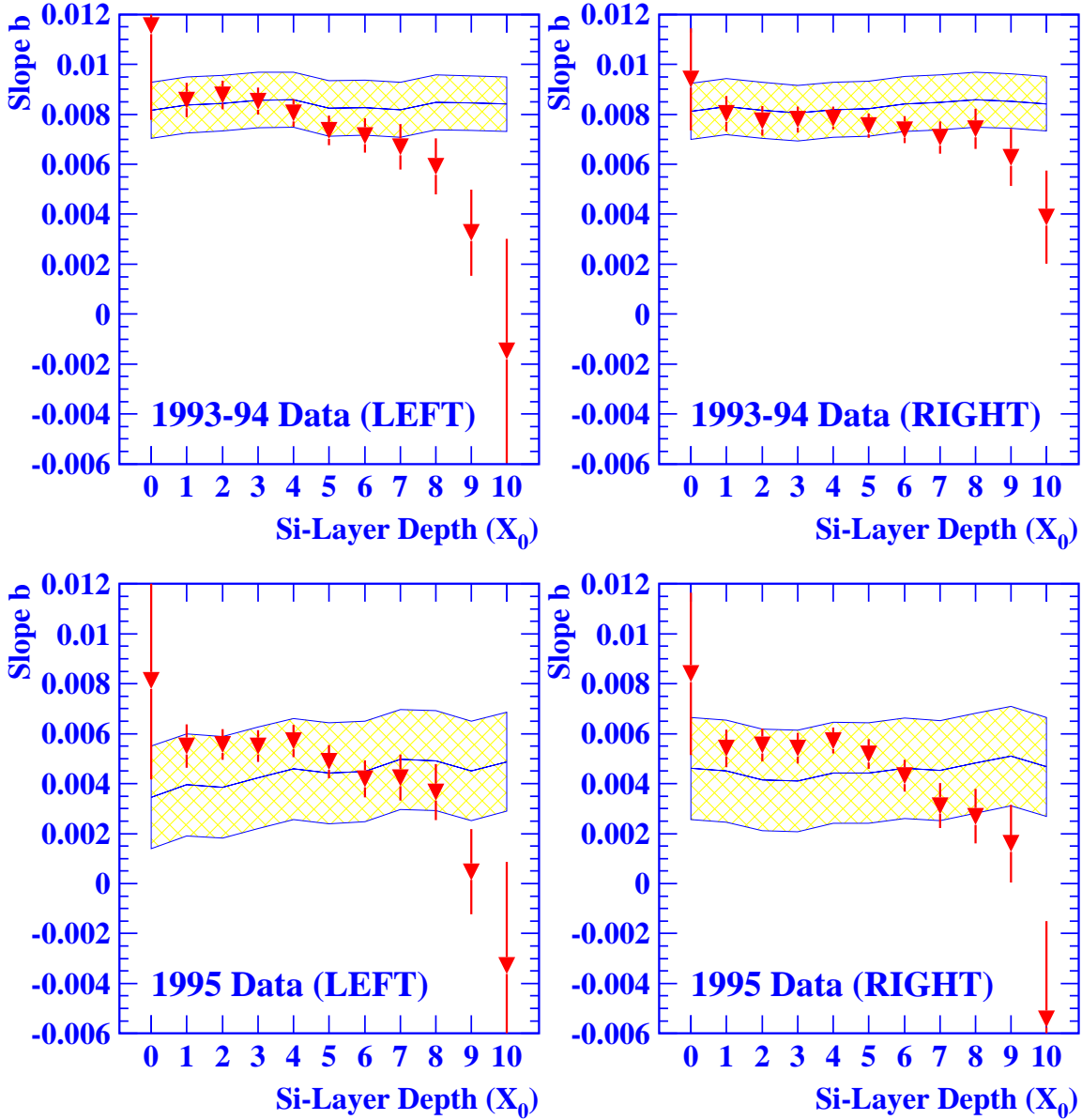}
\caption
{Fitted values of $b$ 
as a function of the anchoring layer for homogeneous data subsamples. 
In each case the band shows the result obtained from the unanchored
radial distribution, where the width corresponds to the statistical error
and the small shifts depend on
the rebinning of the distribution in each layer.
The solid triangles show results from distributions
corrected by anchoring, with the error bars representing 
only the systematic errors.}
\label{fig:b-vs-layer}
\end{center}
\end{figure}

In the second study, we focus on broader systematic effects which have a direct
impact on the measured running.
Here we optimise the number of bins for extraction of the slope $b$,
as specified in Table~\ref{tab:bins} (for layer $4\,\x$).
This binning is more consistent with the small number
of parameters we wish to extract. The bin width grows with increasing
radius to compensate the diminishing statistics, and any noise associated with
fluctuations at the suppressed intermediate bin boundaries, 
which we tested in our first study, is reduced.
Here, in addition to the statistical errors, we also include the systematic
errors of the anchoring procedure, 
determined as described in Section~\ref{sec:ancherr}.
The fitted values of $b$ for each layer are shown in Fig.~\ref{fig:b-vs-layer}.
In each case the band shows the result obtained from the unanchored
radial distribution, where the width corresponds to the statistical error.
Unlike the anchored points, the unanchored results are independent of depth,
and the small shifts only arise from the rebinning
of the distribution in each layer.
By applying the anchoring corrections the slope $b$ does not change more than
its statistical error in layers $1$-$6\,\x$ and then shows a steady
decrease with increasing depth in the calorimeters.
Layer $0\,\x$ shows a strong deviation from the unanchored result, 
in the opposite sense.
For 1993-94 data the onset of deviations on the Left side 
precedes that on the Right by almost exactly one layer,
consistent with Fig.~\ref{fig:dickplot},
due to the presence of the extra preshowering material 
of the microvertex cables.
In the 1995 data the observed deviations are similar on the two sides,
reflecting the fact that the preshowering material is then symmetric.

The anchored slope is remarkably flat
for the broad region covering layers $1$-$6\,\x$ 
in the 1993-94 Right data. 
Here the individual results 
are consistent within the assigned systematic errors.
For the other datasets the values of the anchored slope in layers $1$-$6\,\x$ 
are distributed within twice the assigned systematic errors.
Similarly the difference between the individual values of the anchored 
and unanchored slopes is within twice the assigned systematic 
error in each case.
The Left side appears to exhibit a pattern of deviations
consistent with a residual depth-dependent effect, particularly in the
more statistically significant 1993-94 data.
Due to the back-to-back nature of Bhabha events, the two sides do not
contribute independent statistical information.
We therefore choose to derive our final results from the Right side alone.
Even though in 1995 the Right data are affected by extra preshowering
material to an extent similar to the 1993-94 Left data, we consider that
the gain of statistical precision in using these independent data outweighs 
the risk of any potential increase in systematic error.
Possible unaccounted systematic errors related to the
preshowering material are estimated in section \ref{sec:deadmaterial}.
We also choose the anchors in layer $4\,\x$ to correct our final
results, since this layer lies in the centre of the safe region, where
the anchoring procedure is valid.

\section{Systematic uncertainties}
\label{sec:syst}
Having discussed how we ensure that our anchoring corrections are chosen
from a safe region of the detector, we now quantify the residual errors
we attribute to these anchors, as well as all other systematic errors 
we identify as
affecting the measurement.  The analysis of these errors closely follows
their treatment in the SiW luminosity analysis~\cite{lumipap}.  
The discussion here focuses on those considerations specific to the current 
measurement, and quantifies their effect, $\delta b$,
on the extracted value of $b$. 
The systematic uncertainties are grouped into classes 
and discussed in detail below.

\subsection{Anchoring errors}
\label{sec:ancherr}
This error class includes all the uncertainties connected with the
anchoring procedure described in Section~\ref{sec:anchoring}. 
Biases on the radial coordinate can directly affect the shape of the $t$
distribution. In particular a bias on
the reconstructed position in the central part of the acceptance, behind the
preshowering material, can produce a significant error. As mentioned in
Section~\ref{sec:method}, a bias in the central region of $70 \, \mu$m would
double the expected apparent running or reduce it to zero depending 
on the sign of the bias.

Shifts in the anchors can arise either through the measured value
of $\Roff$ or through the determination of the pad boundary bias, 
$\delta R_{R\phi}$.
The measurement of $\Roff$ is affected by pad gain fluctuations 
and departures from the Gaussian model used for fitting the pad boundary 
image.
A global uncertainty in the pad boundary bias $\delta R_{R\phi}$ has negligible
effect on the current measurement, but uncertainty in its radial dependence
must be considered.

The observed $\Roff$ is affected by 
fluctuations in the individual pad gains.
We have checked these effects directly on data, by studying $\Roff$
for each of the $32$ azimuthal divisions of the calorimeters.
The size of the extra azimuthal variations,
after allowing for the statistical and the metrology fluctuations, 
is taken as the systematic uncertainty due to pad gain variations. 
We assign the statistically expected shift in the mean 
$(\Roff)_{\mathrm{RMS}}^{\mathrm{extra}} / \sqrt{32}$ as a systematic error 
in the anchors.
This is a genuine uncorrelated error as a function of radius, estimated to
contribute $\delta b = 32 (30) \times 10^{-5}$ for the 1993-94 (1995) data. 

Another error component uncorrelated with radius is the statistical error
of the anchors, determined from the fits of the pad boundary images.
This gives a further $\delta b = 10 (20) \times 10^{-5}$
for the 1993-94 (1995) data.

Fig.~\ref{fig:pbimage} shows that a Gaussian resolution does not perfectly
describe the tails of the pad boundary image. To the extent that this
image maintains an odd symmetry about the apparent pad boundary, its
non-Gaussian behaviour does not affect the determination of $\Roff$, 
as can be
seen from the close agreement of the data points and the fitted curve near the
pad boundary. We have also considered a model in which the apparent pad boundary
is taken as the median of the observed resolution function. 
The shift in the slope which is obtained with this alternative method is taken
as an error. In the safe region of the detector it is very small,
$\delta b = 2 (12) \times 10^{-5}$ for 1993-94 (1995) data.
In the deeper layers of the detector, however, this error source
becomes significant, and may be underestimated.  In fact we suspect that this
effect may contribute to the observed run-away of the anchors at large depths.

The determination of the pad boundary bias in the test beam was carried
out at a radial position close to the inner acceptance cut to provide
optimal information for the luminosity determination. In this analysis
we have a greater dependence on knowing the pad boundary bias throughout
the detector.
The geometrical bias due to $R$-$\phi$ pads is expected to scale as $1/R$, 
thus decreasing at a greater radius of pad curvature. Therefore we have
scaled the bias estimated using the test beam results, but assign
an additional systematic error equal to $50$\% of the expected bias to account
for possible deviations from this behaviour.
We have estimated conservatively the effect of this additional error, 
by shifting in a correlated manner all the anchors by this error
and taking the resulting variation in the slope.
This is found to be $\delta b = 25 (32) \times 10^{-5}$
for the 1993-94 (1995) data.

Further errors on the determination of $\delta R_{R\phi}$ have been estimated 
for the worst case of a bias 
anticorrelated between the inner and the outer radius.
Under this conservative hypothesis, the uncertainty on the 
shower width dependence of the pad boundary bias 
may contribute a shift of $\delta b = 11 \times 10^{-5}$, 
common to all the datasets.
The uncertainty on the input apparent width $\sigmaa$ may shift the slope by 
$10 (19) \times 10^{-5}$ for the 1993-94 (1995) data. 

All these error contributions have been summed in quadrature, giving finally
a total anchoring error on the slope $b$ of 
$45 (52) \times 10^{-5}$ for the 1993-94 (1995) data. 
This corresponds to an effective anchoring error of about $4~\mu$m.

\subsection{Preshowering material}
\label{sec:deadmaterial}
To cover in a conservative way possible effects of the preshowering material 
we made a few direct tests using the data.
The amount of preshowering material is maximum in the middle of the accepted
radial range, as is reflected in the $\sigmaa$ 
distribution shown in Fig.~\ref{fig:widths}. We have thus defined two regions:
\begin{itemize}
\item {\it clean} region,
$R \leq 8.2$~cm and $R \geq 11.7$~cm, corresponding to the
first 4 pads starting from the inner radial cut and the last 6 pads close to the
outer cut;
\item {\it obscured} region,
$8.2 < R < 11.7$~cm, corresponding to the central 14 pads. 
\end{itemize}
The fitted slopes determined separately in the two regions and in the full acceptance
are given in Table~\ref{tab:fitclean}.
We see that the results obtained in the clean region are quite 
close to the results of the fit over the full acceptance.
Moreover the independent values obtained in the obscured region 
are consistent within the statistical errors.
It is natural to expect a possible extra pad boundary bias 
in the obscured region, particularly on the Left side. 
We have checked for its presence by introducing a new parameter $x$ in
the fit, related to this assumed extra bias, using two alternative models:
\begin{itemize}
\item {\em Box-model}, 
a naive choice assuming a constant extra bias within the obscured region
and no extra bias in the clean region. Here $x$ is the constant extra bias.
\item {\em W-model}, 
the extra bias $\deltaRextra$ is assumed to follow the pattern of
the apparent shower width $\sigmaa$ versus $R$:
\begin{equation}
\deltaRextra = 
x \frac{\sigmaa(R) - \sigmaa(\Rmin)}{\sigmamax - \sigmaa(\Rmin)}
\end{equation}
where $\Rmin$ is the inner acceptance cut (where $\sigmaa$ is minimum) and
$\sigmamax$ is the maximum value of $\sigmaa$, which is reached near 
the centre
of the detector ($R \approx 10.2$~cm). Therefore $\deltaRextra = x$ when 
$\sigmaa(R) = \sigmamax$. 
\end{itemize}
No evidence for an extra bias is found under either of the two hypotheses. 
We take the statistical sensitivity of the check as a systematic uncertainty, quantified as
the additional contribution to the error on the slope generated by allowing the
possibility that such an extra bias might exist.
We obtain $30 (53) \times 10^{-5}$ for the Right (Left) side in 1993-94 data
and $87 (94) \times 10^{-5}$ for the Right (Left) side in 1995 data.
Notice that such uncertainties cover the observed shifts between fits in the
restricted clean region and in the full acceptance 
in each of the various cases.

\subsection{Position resolution}
The radial resolution at pad boundaries in the clean
acceptance near the inner edge of the detector has been measured using the test
beam to be $130~\mu$m. The apparent resolution at the outer edge and in the
central portion of the detector, behind the bulk of the preshowering material,
is degraded by a factor of 2 to 2.5, according to the pattern of
Fig.~\ref{fig:widths}. The Monte Carlo simulation includes a radial
dependence accounting for this variation.
The impact of any unaccounted degradation of the radial resolution as a function
of radius is tiny. For example, to get an effect the same size as the
running, the resolution behind the material would have to be wrong by 2~mm.
We conservatively 
assessed the uncertainty related to the radial resolution by dividing the
acceptance into two radial bins and calculating the full effect of the 
resolution flow on the slope. 
It amounts to  $15 (25) \times 10^{-5}$ for 1993-94 (1995) data.
The contribution of the resolution flow across the acollinearity cuts is
negligible in comparison, 
amounting to less than $5 \times 10^{-5}$ in all cases.

The resolution of the reconstructed azimuthal coordinate 
is not critical because of the cylindrical symmetry of the detector. 
It only enters through the cuts on the
acoplanarity distribution as a resolution flow effect, which is taken into
account by the detector simulation. 
Radial variation of the azimuthal resolution and unaccounted 
non-Gaussian tails give uncertainties on the slope smaller than 
$3 \times 10^{-5}$ and have been neglected.

\subsection{Acollinearity bias}
In contrast to the radial distribution,
the acollinearity distribution, with the selection cut $|\Delta R| \leq
2.5$~cm, is not corrected for radial biases. 
It is therefore subject
to biases of the order of the anchors themselves. In the worst case there could
be a first order effect causing a net gain or loss of events at both the
positive and the negative $\Delta R$ cut. This is conservatively estimated by
considering a bias with absolute value $\Delta R_{\mathrm{bias}} = 30~\mu$m, 
which is the maximum reached by the anchors. 
This corresponds to an uncertainty on the slope of $9 \times 10^{-5}$.

\subsection{Metrology}
The detector geometry was carefully determined and monitored
for the luminosity measurement \cite{lumipap}. The most crucial quantity in
that analysis was the inner
radius of the calorimeters, since the internal geometry of the Si wafers is inherently
excellent.  This analysis is much less sensitive to the absolute radial scale,
and it would require a radial shift of about $0.5$~mm to mimic the expected $\alpha$ running.
Even without averaging over the measurements of the two sides, 
the inner radii are known with a precision of $7\,\mu$m, 
which gives an error on the slope of only $12 \times 10^{-5}$.

In operating conditions thermal effects also contributed variations to
the nominal radial dimensions of the detector on the order of $2$-$10 \,\mu$m. 
These were calculated independently for each data set, 
according to the average temperature measured by thermistors located on each
detector layer. Such thermal effects give contributions to the
slope of $3$-$13 \times 10^{-5}$, which have been considered
correlated between all the data samples.

This analysis is insensitive to longitudinal metrology errors. In fact an error
in the longitudinal separation between the Right and Left calorimeters would be
equivalent to a shift in the $z$ position of the interaction vertex 
which has no effect on the slope.
Also the inter-layer separation gives negligible uncertainties: 
for a change
of $100\,\mu$m in the spacing between layers, the effect on the slope is about 
$1 \times 10^{-5}$ for all layers considered ($1$-$10 \,\x$).

\subsection{Beam parameters}
The geometry of the colliding beams with respect to the detector can
have quite important effects on the apparent slope.
The expected $\alpha$ running has an effect comparable to an uncorrected 
transverse offset of the beams of 4 mm or
a beam angular divergence of $1.4$~mrad. 
These values are much larger than those we experienced. 
Moreover we were able to determine such parameters from the data.

Variations in the beam parameters modify the geometric acceptance
of the definition cuts in a manner which can be adequately calculated
analytically by assuming a simple $1/\theta^3$ angular distribution.
The effect of the isolation cuts is less tractable analytically, but
since they considerably decrease the variations
calculated from the acceptance cuts alone, their neglect yields
a conservative estimate of the uncertainties.
The effects on the slope $b$ may be estimated with sufficient
precision by dividing the radial acceptance into two bins.

The definition cuts are sensitive only to the sum of the transverse beam offset
and the beam tilt.
The transverse beam offset is measured run-by-run with a precision better than
$10\,\mu$m and gives a negligible uncertainty on the fitted slope.
The beam tilt is therefore the most important effect. 
Its two components are determined run-by-run as the
difference of the eccentricities of the unscattered beams as they pass through
the bore of each calorimeter.
These eccentricities are measured by the azimuthal modulation of the Bhabha
intensity, and the statistical accuracy with which they can be determined
is $200$-$300 \,\mu$m for typical runs. For the nine data sets the beam tilt
contributes $2$-$9 \times 10^{-5}$ to the uncertainty in the slope.
These errors have been conservatively taken
twice: both correlated and uncorrelated.
Although they are unlikely, we have no way of measuring rapid tilt
variations, on a time scale shorter than an individual run, we have therefore
taken as an additional uncertainty the slope variation
corresponding to the extreme hypothesis of setting the beam tilt to zero. 
This uncertainty ranges from $2 \times 10^{-5}$ 
for the largest sample (94-b) to $32 \times 10^{-5}$ in the worst
case. We have taken such numbers as uncorrelated errors, with an additional
common correlated systematic error equal to $2 \times 10^{-5}$.

The transverse beam size and divergence give effects similar to the 
radial resolution and can be estimated, in an analogous way, from the related
resolution flow.
The full effect of the beam size on the slope is about
$1 \times 10^{-5}$, and has been neglected.
The uncertainty on the beam divergence, estimated by comparing two independent
determinations, is about $100\,\mu$rad for 1993-94 data and about
$130~\mu$rad for 1995 data. 
The resulting uncertainties on the slope are
$4 \times 10^{-5}$ and $7 \times 10^{-5}$ respectively. 
These errors have been conservatively taken
twice: both correlated and uncorrelated.

The longitudinal position of the beam spot has a constant effect on the radial
acceptance so gives no contribution to the slope. The same holds for the
longitudinal size of the beam spot.

The uncertainties estimated analytically have been checked with the results of
Monte Carlo simulations where one parameter at a time can be varied. 
In this manner the isolation cuts are included, at the price of some statistical
limitation. The results are consistent.

\subsection{Energy response}
The essential 
problem here is how the preshowering material degrades the energy response 
and causes events to be lost from the accepted sample.
Uncertainties due to the energy response have been assessed 
by varying the parameters in the detector simulation
within the precision to which they have been estimated from the data. 
They include the 
Gaussian width of the energy response function, the exponential low-energy tail,
the nonlinearity and 
the method used to extrapolate the energy resolution to lower energies.
We determined the variation in the fitted slope in a conservative way, by
dividing the radial acceptance into two bins and then by changing in turn
each parameter in the simulation of the outer bin, leaving unchanged the
parameters for the inner bin. Then we took the
sum in quadrature of all the variations.
The dominant uncertainty is caused by the low-energy tail of the
response function, which is generated by events that shower very late in the detector,
events not fully contained and events with electrons and positrons that scatter
in upstream material.
The resulting uncertainty is $27 \times 10^{-5}$.

\subsection{Accidental background}
Off-momentum electrons and positrons generated by beam-gas
scattering generate the majority of single showers observed in the luminometer.
Accidental coincidences between background clusters in the Right and 
Left calorimeters can occasionally produce events which are selected as Bhabha scatterings,
although in general these events are rejected by the minimum energy cuts 
or fail the acollinearity cuts.
This background was studied by special triggers including random bunch crossings
and delayed coincidences. The background fractions were found to 
vary between $0.1$ and $0.6 \times 10^{-4}$ depending on the data set
\cite{lumipap}. 
We have estimated the effect on the slope by dividing the
acceptance into two radial bins and assuming that 
the background is concentrated in the inner bin. 
The radial distribution of background is indeed 
sharply peaked at low radius. To account for the enlarged acceptance used in
this analysis we increased the background fractions by $50$\%, which covers the
variations observed by studies of the background distributions.
The effect on the slope is a slight decrease of between
$2$ and $16 \times 10^{-5}$, for which no correction has been applied.
The full effect is taken as an uncertainty correlated between the data samples.  

The accidental overlap of a background cluster with a Bhabha event can also
change the values of reconstructed quantities, modifying the acceptance.
This has been evaluated by incorporating the measured background into the
detector simulation, by adding to BHLUMI events background clusters
with rates determined from random triggers. 
Likewise we have also combined the measured background with data.
In both cases the resulting shift in the slope is 
smaller than $10^{-5}$ and has been neglected.

\subsection{Physics biases in the fitted slope}
\label{sec:biases}
The fit method used to determine the slope $b$
has been studied on Monte Carlo samples.
The logarithmic $t$ dependence expressed by 
Equation~(\ref{eq:Rfit}) is followed very well. Actually in data we
have no statistical sensitivity to the difference between a logarithmic
and a simple linear behaviour.
However, the parameterization in Equation~(\ref{eq:Rfit}) 
has the advantage that the
logarithmic slope $b$ is almost constant for small variations of the $t$ range
or the different centre-of-mass energies which we have in data.
The fit method has a small bias at off-peak energies, due to 
effects of $\Z$ interference, of about $\delta b = \pm 14 \times 10^{-5}$,
where the plus sign applies to energies above the peak energy
and the minus sign below.
This bias has been corrected on each data subsample, with negligible
uncertainty. 
The final average would not suffer in any case from this small effect,
which cancels when averaged over the
roughly equal samples above and below the peak.

The reconstructed $t$ has a small positive bias, due to initial-state radiation,
with respect to the exchanged four-momentum between the electron and positron
line. 
From a study of BHLUMI Monte Carlo events 
this bias is found to be almost zero at
the inner acceptance cut, increasing to about $0.1$~GeV$^2$ at the outer cut. 
This small variation of the $t$ range has a negligible effect on the slope $b$.

Our selection contains a small irreducible physics background from the process
$\epem \rightarrow \gamma \gamma$, for which we apply a correction. 
Its cross section within our
idealized acceptance is found to be $16.9$~pb at $91.1$~GeV using a Monte Carlo
generator including \cal{O}($\alpha^3$) terms \cite{radcor}.
The correction to
the slope $b$ is $-18 \times 10^{-5}$, practically constant
with respect to our range of centre-of-mass energies, with negligible
uncertainty.

\subsection{Summary of systematic errors}
\label{sec:syssummary}
The dominant systematic errors are those related to anchoring and preshowering
material, described in Sections~\ref{sec:ancherr} and \ref{sec:deadmaterial}.

The experimental systematic uncertainties are summarized in
Table~\ref{tab:syserr_r}, where
the nine data samples are labelled as in \cite{lumipap}.
The final experimental error correlation matrix 
(including the statistical errors)
is given in Table~\ref{tab:corrmat}. The correlations reach at most $10$\%.
The classification of the detailed sources of
error into correlated and uncorrelated components given in 
Table~\ref{tab:syserr_r} does not reveal the complete pattern
of correlations embodied in the full correlation matrix. 
In that table the errors
classified as correlated are fully correlated between all data samples, while
some classified as uncorrelated are actually correlated within a given year, but
uncorrelated between years. 

\section{Theoretical uncertainties}
\label{sec:theory}
It is important to assess the theoretical uncertainties associated with
our reference cross section calculated by the BHLUMI Monte Carlo.
In fact a reliable determination of the running coupling
constant from Equation~(\ref{eq:xsec}) 
needs a precise knowledge of the radiative corrections.

The BHLUMI Monte Carlo is a multiphoton exponentiated generator for small-angle
Bhabha scattering.
Its matrix element includes complete ${\cal O}(\alpha)$ and
${\cal O}(\alpha^2 L^2)$ photonic corrections.
Higher order contributions are partially included 
by virtue of the exponentiation.
The theoretical uncertainty in the BHLUMI calculation has been studied 
extensively for the event selections of the LEP experiments. 
The fractional theoretical error is $6.1 \times 10^{-4}$ for
the integrated cross section at \LEPone~\cite{yellowrep,bflward}
which was relevant for the determination of the luminosity.
Alternative existing calculations have been widely cross-checked with BHLUMI
\cite{yellowrep}. 
Moreover the extensive comparisons between data and the predictions of 
the Monte Carlo by the four LEP collaborations decrease the chances that 
it contains significant residual technical imperfections. 

We also use two other Monte Carlo generators which are included in the same
BHLUMI package. 
OLDBIS \cite{bk83,oldbis} is an exact ${\cal O}(\alpha)$ calculation, 
based on a Monte Carlo program written by an independent group.
LUMLOG \cite{bhlumi,lumlog} implements a Leading-Log calculation up to 
${\cal O}(\alpha^3 L^3)$, based on a structure function approach, 
assuming purely collinear radiation. 
The BHLUMI package gives access to many intermediate weights 
which compose the final calculations, 
so that we have also checked several different approximations.

For the purpose of assessing the theoretical uncertainties, the OPAL
experimental selection has been described by a slightly modified version
of the code used in \cite{yellowrep} and included in the BHLUMI package
(subroutine {\tt TRIOSIW}).
Events have been generated within a safely enlarged angular region to protect
against loss of visible events.
Smearing effects are neglected and an ideal beam geometry is assumed. 
Nearby particles are combined by a clustering algorithm 
which has a window matched to the experimental
resolution. The energy is defined by summing all the particles inside the
isolation cuts in each calorimeter. 
The position variables $R$ and $\phi$ are defined as the
coordinates of the highest energy particle reconstructed on each side.
We have applied all the isolation cuts listed 
in Section~\ref{sec:selection} to these reconstructed variables. 
The size of the window used by the clustering algorithm (in $R$ and $\phi$) has
been varied over a large range to verify the stability of the result.
As further checks we have used alternative selections, for example
SICAL (following the nomenclature of \cite{yellowrep}) which mostly differs in
its lack of acollinearity cuts, and BARE, a non-calorimetric selection.

\begin{figure}
\begin{center}
\epsfxsize=\textwidth
\epsfbox{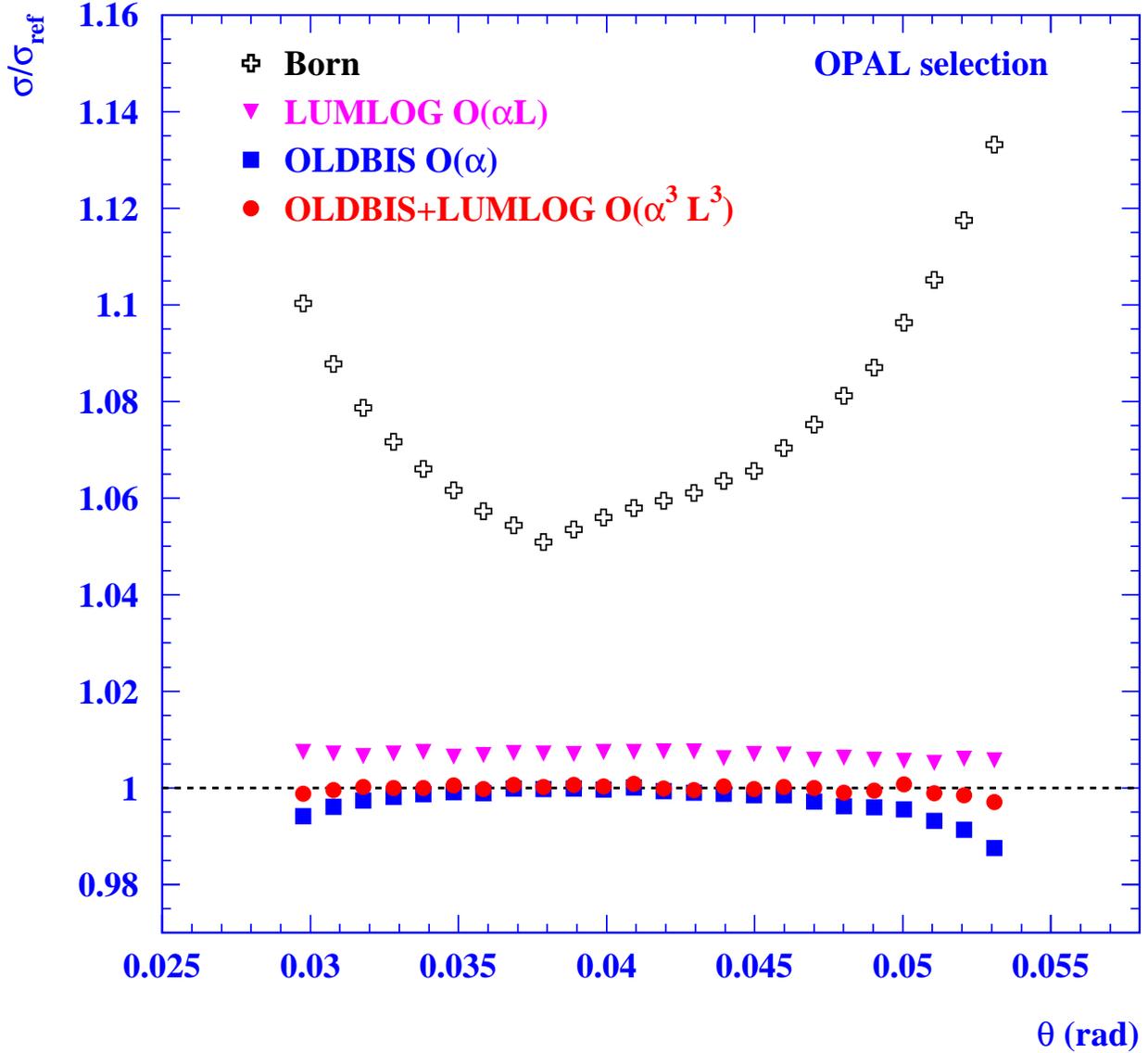}
\caption
{Differential cross section as a function of the polar scattering angle,
in different perturbative approximations 
normalized to the reference BHLUMI calculation, for the OPAL selection.
The reference BHLUMI (${\cal O}(\alpha^2 L^2)$ exponentiated) is
shown as the dashed horizontal line at $\sigma/\sigma_{\mathrm{ref}}=1$.
Vacuum polarization, $\Z$-interference and $s$-channel photon
contributions are switched off.}
\label{fig:diffxsec}
\end{center}
\end{figure}

The differential cross section obtained at
different perturbative orders is shown in Fig.~\ref{fig:diffxsec} normalized
to the reference BHLUMI cross section. Here vacuum polarization, $\Z$-exchange
interference and $s$-channel photon interference have been switched off.
Radiative corrections reduce the Born cross section
by about $5$-$15$\%, depending on the polar
angle. Most of the reduction is obtained by the leading ${\cal O}(\alpha L)$
corrections, which give a result uniformly about $1$\% above the reference.
When the complete ${\cal O}(\alpha)$ corrections are considered,
including subleading terms,
the result lies below, generally within $1$\% of the reference 
but with stronger edge effects.
The most precise alternative calculation shown in Fig.~\ref{fig:diffxsec}
is the sum of the complete ${\cal O}(\alpha)$ given by OLDBIS and the 
leading higher orders (${\cal O}(\alpha^2 L^2)$ and ${\cal O}(\alpha^3 L^3)$)
given by LUMLOG. This combination was previously used in \cite{yellowrep},
and is termed OLDBIS+LUMLOG in what follows.

Examination of the canonical coefficients \cite{pvrncim} indicates that for a
calorimetric detector acceptance such as ours the ${\cal O}(\alpha^2 L)$ and
${\cal O}(\alpha^3 L^3)$ terms dominate the approximately
calculated portion of the complete BHLUMI cross section.

The effect of ${\cal O}(\alpha^3 L^3)$ terms on the slope $b$ measured 
within our selection can be
directly calculated using LUMLOG and is found to be small,
$(-10 \pm 9) \times 10^{-5}$. Moreover such terms are known to be 
almost completely accounted for by the exponentiation in BHLUMI.

The magnitude of the ${\cal O}(\alpha^2 L)$ corrections can be roughly 
estimated from the product of the leading ${\cal O}(\alpha L)$ and 
the subleading ${\cal O}(\alpha)$ corrections.
This is about $0.5 \times 10^{-3}$
times the Born cross section when integrated over the entire acceptance.
The effect on the slope $b$ is estimated by calculating this product for
the inner and outer halves of the acceptance separately and 
then taking the difference. 
In this way we obtain $\delta b = 20 \times 10^{-5}$, which is
expected to be correct within a factor of about $2$. 

\begin{figure}
\begin{center}
\epsfxsize=\textwidth
\epsfbox{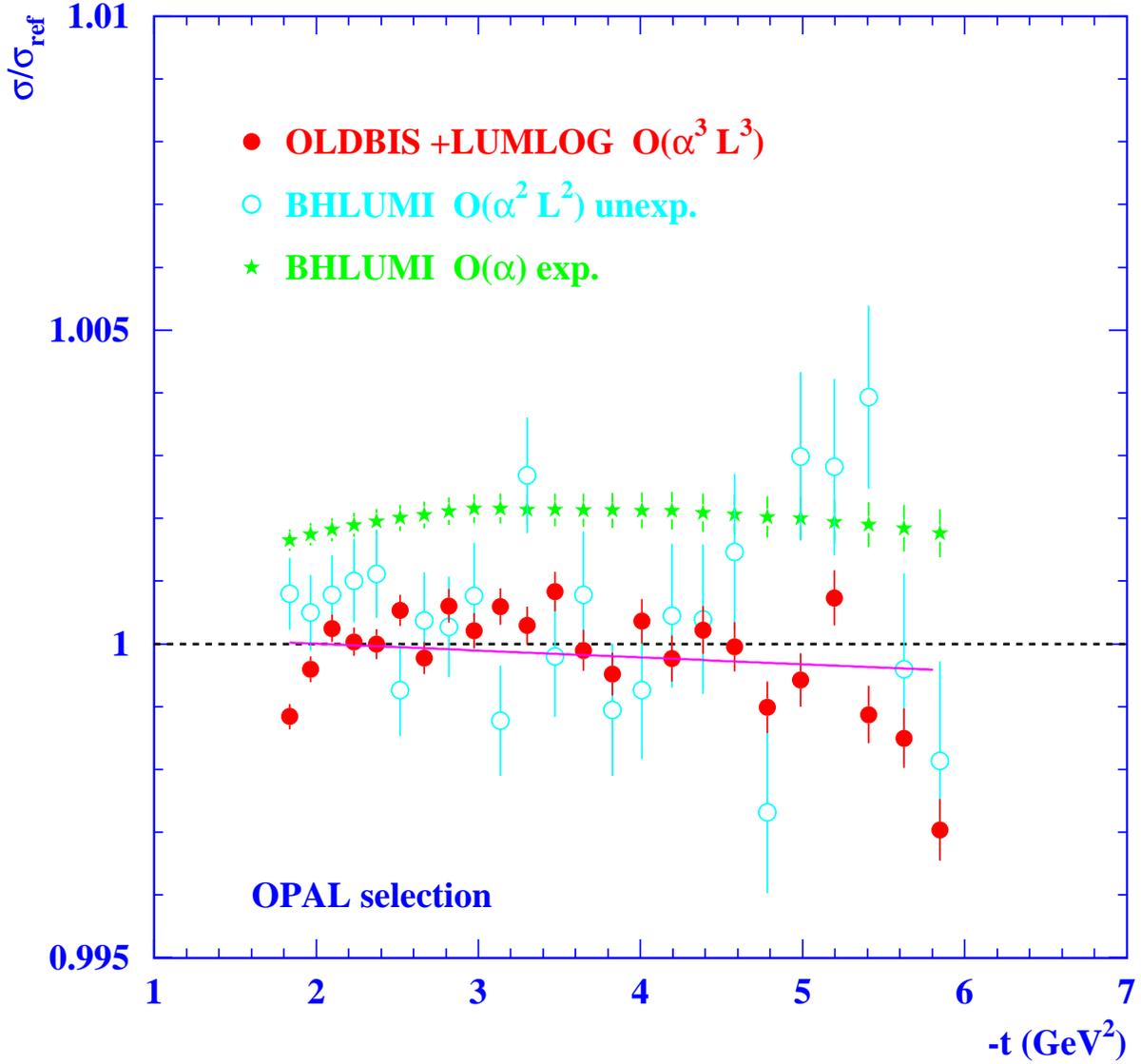}
\caption
{Differential cross section as a function of $-t$,
in different perturbative approximations normalized to the reference BHLUMI
calculation, for the OPAL selection.
The reference BHLUMI (${\cal O}(\alpha^2 L^2)$ exponentiated) is
shown as the dashed horizontal line at $\sigma/\sigma_{\mathrm{ref}}=1$. 
Vacuum polarization, $\Z$-interference and $s$-channel photon exchange
contributions are switched off.
The superimposed solid line is a fit to the OLDBIS+LUMLOG result.}
\label{fig:obilog}
\end{center}
\end{figure}

A better test of the ${\cal O}(\alpha^2 L)$ corrections, which
also takes into account the limited technical precision of
the Monte Carlos, can be obtained by comparing the reference 
${\cal O}(\alpha^2 L^2)$ exponentiated calculation of BHLUMI to the 
combination of the two independent Monte Carlos, OLDBIS and LUMLOG.
The OLDBIS+LUMLOG combination is shown together with
other precise calculations at our disposal
in Fig.~\ref{fig:obilog}. 
To quantify possible deviations from the expected $t$ shape,
we fit the ratios of alternative calculations to the reference BHLUMI 
with a linear $t$ dependence.
The effect, $\delta b$, on our measured slope $b$ 
is then estimated by accounting for the
factor $|t_2-t_1| / \ln(t_2/t_1) = 3.52$ which converts from a linear to a
logarithmic $t$ range.
We obtain:
\begin{eqnarray}
\delta b {\, [ {\rm (OLDBIS+LUMLOG)} / {\rm BHLUMI} ]} =
(-38 \pm 20) \times 10^{-5} \,\phantom{.}
\nonumber \\
\delta b {\, [ {\rm Exp.}{\cal O}(\alpha) / {\rm BHLUMI} ]} =
(+22 \pm 16) \times 10^{-5} \,\phantom{.}
\nonumber \\
\delta b {\, [ {\rm Unexp.}{\cal O}(\alpha^2 L^2) / {\rm BHLUMI} ]} = 
(-11 \pm 61) \times 10^{-5} \,.
\nonumber
\end{eqnarray}
The ratio of the exponentiated ${\cal O}(\alpha)$ calculation to the full BHLUMI
is quite flat as a function of $t$, with a clear normalization shift of $0.2$\%,
which is however irrelevant to our analysis. 
The unexponentiated ${\cal O}(\alpha^2 L^2)$ also seems flat, albeit with
much larger statistical error bars. 
The OLDBIS+LUMLOG combination is rather flat with
somewhat larger than expected fluctuations. The extreme points on either end
show downward deviations with respect to the reference BHLUMI, but removing them
does not change the fit significantly. 
We take the shift $\delta b$ determined from the
${\rm (OLDBIS+LUMLOG)} / {\rm BHLUMI}$ fit
as the uncertainty due to missing higher orders, which are
mainly ${\cal O}(\alpha^2 L)$, plus the technical precision of the calculations.
This amounts to $38 \times 10^{-5}$, which is in agreement
with our cruder estimate obtained from the product
of the leading ${\cal O}(\alpha L)$ and 
the subleading ${\cal O}(\alpha)$ terms, above.
This estimate is also seen to be in line with the differences observed for
the exponentiated ${\cal O (\alpha)}$ and 
the unexponentiated ${\cal O}(\alpha^2 L^2)$ calculations.
Changing the size of the window used by the clustering algorithm
produces negligible variations.  Similar results are also obtained for
the SICAL selection.

The interference with the $\Z$-exchange amplitude in the $s$-channel is a small
correction, designated $\deltaZ$ in Equation~(\ref{eq:xsec}), which is
not factorized with respect to the main contribution and 
the running coupling constant. 
It is energy dependent,
vanishing at $\sqrt{s} = m_\Z$ and changing sign across the $\Z$ pole. 
In BHLUMI it is calculated up to exact ${\cal O}(\alpha)$ 
photonic corrections, which can also be exponentiated \cite{zinter}. 
Vacuum polarization can be included.
Event samples have been generated at three different energies: 
the $\Z$-peak energy ($\sqrt{s} = 91.1$~GeV) and energies offset by
$\pm 2$~GeV.
At each energy we consider independently the shifts in the slope $b$ 
produced by switching off the exponentiation 
or the vacuum polarization for the calculated interference term $\deltaZ$, 
and then add these shifts in quadrature. 
The maximum value of $30 \times 10^{-5}$, which is obtained at the peak,
is taken as the uncertainty due to $\Z$ interference. 

Concerning the contribution of the vacuum polarization to the $\deltaZ$ term,
this subtle effect could in principle perturb the asserted cleanliness 
of the measurement.
However such an effect is vanishingly small at the peak energy
( $\delta b = -6 \times 10^{-5}$ at $\sqrt{s} = 91.1$~GeV)
and has approximately equal and opposite values smaller
than $\pm 20 \times 10^{-5}$ above and below the pole.
The overall effect is therefore negligible.

BHLUMI does not include diagrams with extra light lepton pairs 
($\epem$, $\mu^+\mu^-$). Their contribution was calculated explicitly 
for the OPAL selection, giving a fractional correction of 
$(-4.4 \pm 1.4) \times 10^{-4}$ \cite{pavia} on the integrated cross section. 
The leading order contribution can be checked with LUMLOG, and gives effects on
the slope $b$ below $11 \times 10^{-5}$  
with the OPAL or the SICAL data selections.

The estimated theoretical uncertainties are summarized in Table~\ref{tab:thunc}.
Their quadratic sum is $50 \times 10^{-5}$ and will be added to the experimental
errors.

\section{Results}
\label{sec:results}
The final results are based on the 
radial distribution observed in the Right side calorimeter,
corrected by the anchoring procedure, 
as explained in Sections~\ref{sec:anchoring}-\ref{sec:fit}.
The correction employs the anchors determined in layer $4\,\x$, 
which is in the middle of the safe region. 
Within this region the anchored and unanchored results have been shown 
to be in fairly good agreement.
The decision to use only the Right calorimeter and to use anchored coordinates,
as well as the choice of the anchoring layer, 
are aimed to achieve our best experimental accuracy and to
minimize possible unassessed systematic errors. 
Consistent results, within the quoted final systematic error,
are in fact obtained if any of these choices are changed.

The radial distribution is binned as specified in Table~\ref{tab:bins}.
The numbers of data and Monte Carlo events in each bin
for the largest subsample (94b) are reported in Table~\ref{tab:results.94b}.
Note that here the Monte Carlo assumes $\alpha(t) \equiv \alpha_0$.
The bin-by-bin acceptance corrections, 
obtained from Equations~(\ref{eq:anc1}) and (\ref{eq:anc2})
by inserting the estimated radial biases at the relevant bin boundaries,
are also given.

The ratio of data to Monte Carlo is fitted with the logarithmic $t$-dependence
of Equation~(\ref{eq:Rfit}) separately for each dataset and 
the results are reported in Table~\ref{tab:fit.bello}. 
Both the dominant statistical errors and the experimental systematic errors,
which are determined as described in Section~\ref{sec:syst}, are shown.
The small corrections for the irreducible background and $\Z$ interference
have been applied to the quoted $b$ values. 
The nine subsamples (labelled as in \cite{lumipap}) give consistent results, 
with $\chi^2$/d.o.f~$ = 6.9 / 8$ for the average $b$ considering only
statistical errors.
We have checked for a possible residual energy dependence
of the measured slope $b$, which might be observed if the calculated 
$\Z$ interference contribution disagreed with the data.
Within the statistical errors of the data we find no such energy dependence.

We combine the results of the nine subsamples using the
error correlation matrix in Table~\ref{tab:corrmat}, obtaining:
\begin{displaymath}
b = (726 \pm 96 \pm 70 \pm 50) \times 10^{-5}
\end{displaymath}
where here, and also in the results quoted below,
the first error is statistical, the second is the experimental systematic and 
the third is the theoretical uncertainty, discussed in section \ref{sec:theory}.
The total significance of the measurement compared to the hypothesis of $b=0$
is $5.6 \,\sigma$. 

\begin{figure}
\begin{center}
\epsfxsize=\textwidth
\epsfbox{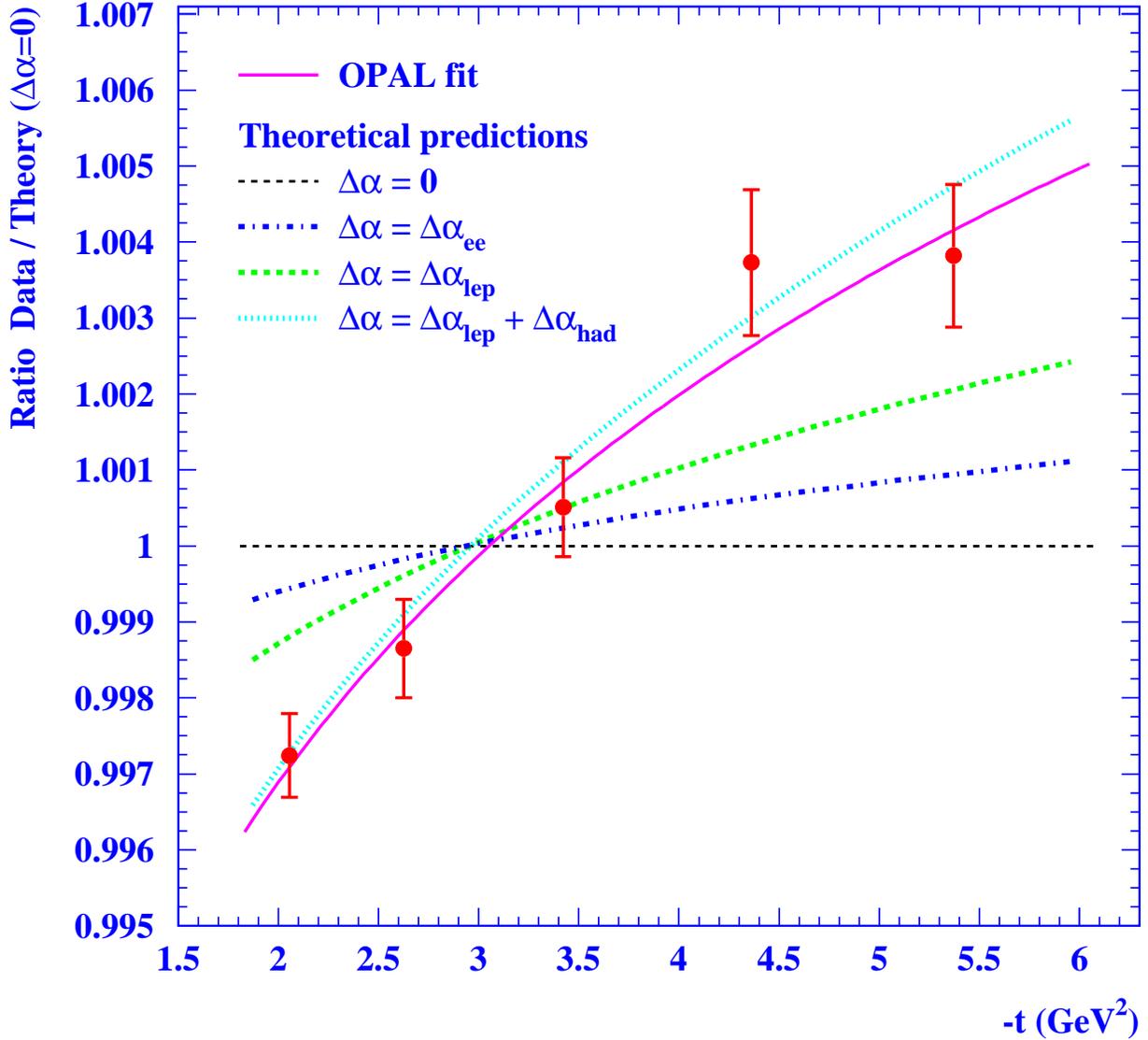}
\caption
{$|t|$ spectrum normalized to the BHLUMI theoretical prediction
for a fixed coupling ($\Delta\alpha = 0$).
The points show the combined OPAL data with statistical error bars.
The solid line is our fit. 
The horizontal line (Ratio=1) is the prediction if $\alpha$ were fixed. 
The dot-dashed curve is the prediction of running $\alpha$ 
determined by vacuum polarization with only virtual $\epem$ pairs, 
the dashed curve includes all charged lepton pairs and 
the dotted curve the full Standard Model
prediction, with both lepton and quark pairs.}
\label{fig:bella}
\end{center}
\end{figure}

The result for the combined data sample is illustrated in Fig.~\ref{fig:bella}.
The error bars shown are statistical only, since many of the systematic errors
are estimated only for the fitted slope in the individual subsamples and
not bin-by-bin.
The logarithmic fit to Equation~(\ref{eq:Rfit}) describes the data very well,
$\chi^2$/d.o.f~$ = 1.9 / 3$,
although a simple linear fit would also be adequate, giving
$\chi^2$/d.o.f~$ = 2.7/ 3$.
The data are clearly incompatible with the hypothesis of a fixed coupling. 
The fitted logarithmic dependence agrees well 
with the full Standard Model prediction including both leptonic and
hadronic contributions, with the hadronic part obtained by the 
Burkhardt-Pietrzyk parameterization \cite{bp2001}.
Older parameterizations of the hadronic component,
like \cite{ej95} or \cite{bp95}, 
would be indistinguishable from the curve shown.
Our result is consistent with the similar measurement by L3 \cite{l3}. If the
latter is expressed as a slope according to (\ref{eq:beff}), it gives
$b^{(L3)} = (1044 \pm 348) \times 10^{-5}$,
where the error is dominated by experimental systematics.

Our measurement of the effective slope gives the variation of the coupling 
$\alpha(t)$ from Equation~(\ref{eq:beff}):
\begin{displaymath}
\Delta\alpha(-6.07 \,\mathrm{GeV^2}) - \Delta\alpha(-1.81 \,\mathrm{GeV^2}) = 
(440 \pm 58 \pm 43 \pm 30) \times 10^{-5} \,.
\end{displaymath}
This is currently the most significant direct measurement where the
running $\alpha(t)$ is probed differentially within the measured $t$
range.
The result is in good agreement with the Standard Model prediction, 
which gives $\delta\left(\Delta\alpha\right) = (460 \pm 8) \times 10^{-5}$
for the same $t$ interval. Here the error has been calculated by assuming 
the uncertainties on $\Delta\alpha_{had}$ at $t = -1.81 \,\mathrm{GeV^2}$
and $t = -6.07\,\mathrm{GeV^2}$ (of $2.5\,$\% and $2.7\,$\% respectively 
\cite{bolek}), which arise from the hadronic component, are fully correlated. 
If the errors on $\Delta\alpha_{had}$ at the extreme values of $t$ 
are assumed to be uncorrelated, the error would increase to 
$23 \times 10^{-5}$. 

The absolute value of $\Delta\alpha$ in our range of $t$
is expected to be dominated by $\epem$ pairs, with the relevant fermion species
contributing in the approximate proportions:
$\mathrm{e} : \mu : \mathrm{hadron} \simeq 4 : 1 : 2$.
Our measurement is sensitive, however, not to the absolute value of
$\Delta\alpha$, but only to its variation within our $t$ range.
Contributions to the slope $b$ in this range are predicted to be
in the proportion: $\mathrm{e} : \mu : \mathrm{hadron} \simeq 1 : 1 : 2.5$.
Fig.~\ref{fig:bella} shows these expectations graphically.
We can discard the hypothesis of running due only to virtual $\epem$ pairs with
a significance of $4.4\,\sigma$.

The data are also incompatible with the hypothesis of running due only to
leptons.
If we subtract the precisely calculable theoretical prediction for all
leptonic contributions,
$\delta(\Delta\alpha_{\mathrm{lep}}) = 202 \times 10^{-5}$, from the measured result, 
we can determine the hadronic contribution as:
\begin{displaymath}
  \Delta\alpha_{\mathrm{had}}(-6.07 \,\mathrm{GeV^2})
- \Delta\alpha_{\mathrm{had}}(-1.81 \,\mathrm{GeV^2}) = 
(237 \pm 58 \pm 43 \pm 30) \times 10^{-5} \,.
\end{displaymath}
This differs from zero by $3.0\,\sigma$, considering all the errors.
In comparison to other existing direct measurements of the QED coupling,
our result has sufficient sensitivity to expose clearly the 
hadronic contribution to the observed variation of $\alpha$ 
as a function of the momentum transfer. 

\section{Conclusions}
\label{sec:conclusions}
We have measured the scale dependence of the effective QED coupling $\alpha(t)$
from the angular distribution of small-angle Bhabha scattering
using the precise OPAL Silicon-Tungsten calorimeters.
Despite the narrow accessible $t$ range, the method has high
sensitivity due to the large statistics and excellent purity of the data sample.
The challenging aspect of the analysis is controlling the residual bias in the
reconstructed radial coordinate of Bhabha electrons in the
detector to a level below $\approx\!10\,\mu$m 
uniformly throughout the acceptance.
From a theoretical point of view the measurement is almost ideal.
For this kinematic range the process is
almost purely QED, $\Z$ interference is very small and the dominant diagram
is $t$-channel single-photon exchange, while $s$-channel photon exchange is
negligible. Due to its utility in determining the LEP luminosity,
small-angle Bhabha scattering is one of the most precisely
calculated processes at these energies. 
We verified that there is no significant disturbance from 
theoretical uncertainties.

We determined the effective slope of the Bhabha momentum transfer
distribution which is simply related to the average derivative of 
$\Delta\alpha$ as a function of $\ln t$ 
in the range $2$~GeV$^2 \leq -t \leq 6$~GeV$^2$.
The observed $t$-spectrum 
is in good agreement with Standard Model predictions.
We find:
\begin{displaymath}
\Delta\alpha(-6.07 \,\mathrm{GeV^2}) - \Delta\alpha(-1.81 \,\mathrm{GeV^2}) = 
(440 \pm 58 \pm 43 \pm 30) \times 10^{-5} \,,
\end{displaymath}
where the first error is statistical, the second is the experimental systematic
and the third is the theoretical uncertainty.

This measurement is one of only a very few experimental tests of the running of
$\alpha(t)$ in the space-like region, where $\Delta\alpha$ has a smooth
behaviour. 
We obtain the strongest direct evidence for the running of the QED coupling 
ever achieved differentially in a single experiment, 
with a significance above $5\,\sigma$.
Moreover we report clear experimental evidence for the hadronic contribution
to the running in the space-like region, with a significance of $3\,\sigma$.

\section*{Acknowledgements}
We thank M.~Caffo and L.~Trentadue for useful discussions.
We particularly wish to thank the SL Division for the efficient operation
of the LEP accelerator at all energies
 and for their close cooperation with
our experimental group.  In addition to the support staff at our own
institutions we are pleased to acknowledge the  \\
Department of Energy, USA, \\
National Science Foundation, USA, \\
Particle Physics and Astronomy Research Council, UK, \\
Natural Sciences and Engineering Research Council, Canada, \\
Israel Science Foundation, administered by the Israel
Academy of Science and Humanities, \\
Benoziyo Center for High Energy Physics,\\
Japanese Ministry of Education, Culture, Sports, Science and
Technology (MEXT) and a grant under the MEXT International
Science Research Program,\\
Japanese Society for the Promotion of Science (JSPS),\\
German Israeli Bi-national Science Foundation (GIF), \\
Bundesministerium f\"ur Bildung und Forschung, Germany, \\
National Research Council of Canada, \\
Hungarian Foundation for Scientific Research, OTKA T-038240, 
and T-042864,\\
The NWO/NATO Fund for Scientific Research, the Netherlands.
%

\normalsize
\vspace{-0.2cm}
\begin{table}[htbp]
\begin{center}
\begin{tabular}{|l||c|c||c|c||c|}
\hline
Bin   & $\Rin$ & $\Rout$ & $-\tin$  & $-\tout$ &  $<-t>$ \\
\cline{2-3}
\cline{4-6}
      & \multicolumn{2}{c||} {\small{(cm)}}&
        \multicolumn{3}{c|} {\small{(GeV$^2$)}} \\
\hline
1   & 7.2584 & 8.2665   & 1.81 & 2.35 & 2.05 \\
2   & 8.2665 & 9.2746   & 2.35 & 2.95 & 2.63 \\
3   & 9.2746 & 10.7868  & 2.95 & 3.99 & 3.42 \\
4   &10.7868 & 11.7949  & 3.99 & 4.77 & 4.36 \\
5   &11.7949 & 13.3071  & 4.77 & 6.07 & 5.37 \\
\hline
\end{tabular}
\end{center}
\caption[]
{Bin definitions for the radial distribution.
The corresponding $-t$ values are determined assuming a reference
energy $\sqrt{s}=91.2208$~GeV.}
\label{tab:bins}
\end{table}
\vspace{-0.2cm}
\begin{table}[htbp]
\begin{center}
\begin{tabular}{|l||c|c|c|}
\hline
Detector side - Radial range & 1993-94    & 1995     & All  \\
\hline \hline
Right - full acceptance&$757 \pm 109$ & $470 \pm 198$ & $690 \pm \phz95$ \\
\hline
Right - clean   & $739 \pm 129$ & $566 \pm 236$ & $699 \pm 113$ \\
Right - obscured& $778 \pm 239$ & $247 \pm 436$ & $655 \pm 210$ \\
\hline \hline
Left - full acceptance& $685 \pm 108$ & $417 \pm 198$ & $623 \pm \phz95$ \\
\hline 
Left - clean    & $680 \pm 129$ & $417 \pm 236$ & $620 \pm 113$ \\
Left - obscured & $766 \pm 239$ & $414 \pm 437$ & $685 \pm 210$ \\
\hline
\end{tabular}
\end{center}
\caption[]
{Fitted slope for the full, clean and obscured 
radial range of acceptance in units
of $10^{-5}$, for homogeneous data subsamples and all data. 
Both the Right and the Left side results are given.
The errors are statistical only.}
\label{tab:fitclean}
\end{table}
\begin{table}[htbp]
{\footnotesize
\begin{center}
\begin{tabular}{|l||*{9}{r|}}
\hline
 Uncertainty           
& 
\multicolumn{1}{|c|}{ 93 $-2$}  &
\multicolumn{1}{|c|}{ 93 pk}    &
\multicolumn{1}{|c|}{ 93 $+2$}  & 
\multicolumn{1}{|c|}{ 94 a}      & 
\multicolumn{1}{|c|}{ 94 b}      &
\multicolumn{1}{|c|}{ 94 c}      &
\multicolumn{1}{|c|}{ 95 $-2$}  &
\multicolumn{1}{|c|}{ 95 pk}    &
\multicolumn{1}{|c|}{ 95 $+2$}  \\
\hline
\hline
   M.C. Statistics    &&&&&&&&&\\
 ~~~~~~uncorrelated   & 56. & 56. & 56. & 56. & 28. & 103.& 56. & 56. & 56. \\
 ~~~~~~~~correlated   & 0.  & 0.  & 0.  & 0.  & 0.  & 0.  & 0.  & 0.  & 0.  \\
        Anchoring     &&&&&&&&&\\
 ~~~~~~uncorrelated   & 10. & 10. & 10. & 10. & 10. & 10. & 29. & 29. & 29. \\
 ~~~~~~~~correlated   & 44. & 44. & 44. & 44. & 44. & 44. & 44. & 44. & 44. \\
Preshowering Material &&&&&&&&&\\
 ~~~~~~uncorrelated   & 0.  & 0.  & 0.  & 0.  & 0.  & 0.  & 81. & 81. & 81. \\
 ~~~~~~~~correlated   & 30. & 30. & 30. & 30. & 30. & 30. & 30. & 30. & 30. \\
       Radial Resolution &&&&&&&&&\\
 ~~~~~~uncorrelated   & 0.  & 0.  & 0.  & 0.  & 0.  & 0.  & 0.  & 0.  & 0.  \\
 ~~~~~~~~correlated   & 15. & 15. & 15. & 15. & 15. & 15. & 25. & 25. & 25. \\
       Acollinearity Bias &&&&&&&&&\\
 ~~~~~~uncorrelated   & 0.  & 0.  & 0.  & 0.  & 0.  & 0.  & 0.  & 0.  & 0.  \\
 ~~~~~~~~correlated   & 9.  & 9.  & 9.  & 9.  & 9.  & 9.  & 9.  & 9.  & 9.  \\
        Radial Metrology  &&&&&&&&&\\
 ~~~~~~uncorrelated   & 0.  & 0.  & 0.  & 0.  & 0.  & 0.  & 0.  & 0.  & 0.  \\
 ~~~~~~~~correlated   & 12. & 12. & 12. & 12. & 12. & 12. & 12. & 12. & 12. \\
	 Radial Thermal   &&&&&&&&&\\
 ~~~~~~uncorrelated   & 0.  & 0.  & 0.  & 0.  & 0.  & 0.  & 0.  & 0.  & 0.  \\
 ~~~~~~~~correlated   & 3.  & 3.  & 3.  & 4.  & 4.  & 4.  & 12. & 12. & 12. \\
         Beam Parameters &&&&&&&&&\\
 ~~~~~~uncorrelated   & 19. & 31. & 20. & 8.  & 5.  & 12. & 12. & 25. & 33. \\
 ~~~~~~~~correlated   & 7.  & 7.  & 7.  & 7.  & 5.  & 9.  & 8.  & 8.  & 8.  \\
         Energy       &&&&&&&&&\\
 ~~~~~~uncorrelated   & 0.  & 0.  & 0.  & 0.  & 0.  & 0.  & 0.  & 0.  & 0.  \\
 ~~~~~~~~correlated   & 27. & 27. & 27. & 27. & 27. & 27. & 27. & 27. & 27. \\
	 Background   &&&&&&&&&\\
 ~~~~~~uncorrelated   & 0.  & 0.  & 0.  & 0.  & 0.  & 0.  & 0.  & 0.  & 0.  \\
 ~~~~~~~~correlated   & 16. & 12. & 7.  & 4.  & 2.  & 5.  & 4.  & 2.  & 2.  \\
\hline
              Sum      &&&&&&&&&\\
 ~~~~~~uncorrelated    & 60. & 65. & 61. & 58. & 30. & 104.& 104.& 106.& 108.\\
 ~~~~~~~~correlated    & 66. & 65. & 64. & 64. & 64. & 65. & 68. & 68. & 68. \\
\hline
&&&&&&&&&\\
 Total Systematic Error& 89. & 92. & 89. & 86. & 71. & 122.& 124.& 126.& 128. \\
\hline
\end{tabular}
\end{center}
\caption[]
{Summary of the experimental systematic uncertainties on the measurement of the effective
slope $b$ for the nine data sets on the Right side. They are broken down into the
components correlated and uncorrelated among the data sets. All errors
are in units of $10^{-5}$.}
\label{tab:syserr_r}
}
\end{table} 
\begin{table}[htbp]
{\footnotesize
\begin{center}
\begin{tabular}{|l||ccccccccc|}
\hline
Sample & 93 $-2$& 93 pk& 93 $+2$& 94 a& 94 b& 94 c& 95 $-2$& 95 pk& 95 $+2$ \\
\hline \hline
93 $-2$ & 1.00 & 0.04 & 0.04 & 0.04 & 0.07 & 0.02 & 0.04 & 0.03 & 0.04 \\
93 pk   & 0.04 & 1.00 & 0.04 & 0.04 & 0.07 & 0.02 & 0.04 & 0.03 & 0.04 \\
93 $+2$ & 0.04 & 0.04 & 1.00 & 0.04 & 0.07 & 0.02 & 0.04 & 0.03 & 0.04 \\
94 a    & 0.04 & 0.04 & 0.04 & 1.00 & 0.07 & 0.02 & 0.04 & 0.03 & 0.04 \\
94 b    & 0.07 & 0.07 & 0.07 & 0.07 & 1.00 & 0.04 & 0.07 & 0.06 & 0.07 \\
94 c    & 0.02 & 0.02 & 0.02 & 0.02 & 0.04 & 1.00 & 0.02 & 0.02 & 0.02 \\
95 $-2$ & 0.04 & 0.04 & 0.04 & 0.04 & 0.07 & 0.02 & 1.00 & 0.08 & 0.10 \\
95 pk   & 0.03 & 0.03 & 0.03 & 0.03 & 0.06 & 0.02 & 0.08 & 1.00 & 0.08 \\
95 $+2$ & 0.04 & 0.04 & 0.04 & 0.04 & 0.07 & 0.02 & 0.10 & 0.08 & 1.00 \\
\hline
\end{tabular}
\end{center}
\caption[]
{The experimental correlation matrix for the nine data sets, considering both
the statistical and the systematic errors.}
\label{tab:corrmat}
}
\end{table}
\begin{table}[htbp]
\begin{center}
\begin{tabular}{|l||c|}
\hline
Error source      &   $\delta b  \,\, (\times 10^{-5})$  \\
\hline \hline

Photonic corrections      & 38. \\
$\Z$ interference         & 30. \\
Light pairs               & 11. \\
\hline
Total                     & 50. \\
\hline
\end{tabular}
\end{center}
\caption[]
{Estimated theoretical uncertainties on the effective slope $b$.}
\label{tab:thunc}
\end{table}
\begin{table}[htbp]
\begin{center}
\begin{tabular}{|c|r|r|c|c|}
\hline
     &  	  &	&\multicolumn{2}{c|} {Ratio $f = \Ndata/\NMCnorun$}\\
\cline{4-5}
Bin &$\Ndata$&$\NMCnorun$ & anchoring & corrected \\
		     & &  &correction & value     \\ 
\hline      
1 & 1310496 & 1314605 & $-0.00019$ & $0.99668 \pm 0.00087$  \\
2 &  927931 &  930603 & $+0.00082$ & $0.99795 \pm 0.00104$  \\
3 &  941500 &  938601 & $-0.00181$ & $1.00128 \pm 0.00103$  \\
4 &  431654 &  430355 & $+0.00116$ & $1.00418 \pm 0.00153$  \\
5 &  458295 &  455712 & $-0.00032$ & $1.00534 \pm 0.00149$  \\
\hline
\end{tabular}
\end{center}
\caption[]
{Detailed inputs to the fit for the largest data subsample (94b). 
Each row corresponds to a radial bin and
gives the number of data events and the number expected from Monte Carlo 
(normalized to the total number of data events),
the applied anchoring correction, and
the corrected ratio with its statistical error.}
\label{tab:results.94b}
\end{table}
\begin{table}[htbp]
\begin{center}
\begin{tabular}{|l|c|r|c|}
\hline
Dataset & $\sqrt{s}$ & Number    & slope $b$\\
        &    (GeV)   & of events & $(\times 10^{-5})$ \\ 
\hline \hline
  93 $-2$   & 89.4510 &  879549 & $ 662 \pm 326 \pm \phz89$ \\
  93 pk     & 91.2228 &  894206 & $ 670 \pm 324 \pm \phz92$ \\
  93 $+2$   & 93.0362 &  852106 & $ 640 \pm 332 \pm \phz89$ \\
  94 a      & 91.2354 &  885606 & $ 559 \pm 326 \pm \phz86$ \\
  94 b      & 91.2170 & 4069876 & $ 936 \pm 152 \pm \phz71$ \\
  94 c      & 91.2436 &  288813 & $ \phz62  \pm 570 \pm 122$ \\
  95 $-2$   & 89.4416 &  890248 & $ 839 \pm 325 \pm 124$ \\
  95 pk     & 91.2860 &  581111 & $ 727 \pm 402 \pm 126$ \\
  95 $+2$   & 92.9720 &  885837 & $ 156 \pm 325 \pm 128$ \\
\hline \hline
  Average   & 91.2208 & 10227352 & $ 726 \pm \phz96 \pm \phz70$ \\
\hline
\multicolumn{3}{|l} {$\chi^2/$d.o.f. (stat.)}       &  $ 6.9/8$  \\
\multicolumn{3}{|l} {$\chi^2/$d.o.f. (stat.+syst.)} &  $ 6.5/8$   \\
\hline
\end{tabular}
\end{center}
\caption[]
{Fit result for each data subsample and average. 
Each row gives the average centre-of-mass
energy, the number of events and the fitted slope $b$, with the first error
statistical and the second the full experimental systematic.
The average is obtained by using the error correlation matrix given in 
Table~\ref{tab:corrmat}. The $\chi^2$ of the average is given both considering
statistical errors only and for the full error matrix.
}
\label{tab:fit.bello}
\end{table}
%
\end{document}